\documentclass[aps,prl,twocolumn,superscriptaddress,amsmath,longbibliography]{revtex4-2}
\usepackage{graphicx,color,hyperref,xcolor}
\usepackage{soul}
\usepackage{amsfonts}
\usepackage{amssymb}
\usepackage{amsmath}
\usepackage{amsthm}
\usepackage{bm}
\usepackage{braket}
\usepackage{enumerate}

\usepackage{CJKutf8}

\begin{document}
\begin{CJK}{UTF8}{gbsn}

\title{Collective optical properties of moir\'e excitons}

\author{Tsung-Sheng Huang}
\affiliation{Joint Quantum Institute, University of Maryland, College Park, MD 20742, USA}

\author{Yu-Xin Wang (王语馨)}
\affiliation{Joint Center for Quantum Information and Computer Science, University of Maryland, College Park, MD 20742, USA}

\author{Yan-Qi Wang}
\affiliation{Joint Quantum Institute, University of Maryland, College Park, MD 20742, USA}

\author{Darrick Chang}
\affiliation{
ICFO-Institut de Ciencies Fotoniques, The Barcelona Institute of Science and Technology,
08860 Castelldefels (Barcelona), Spain}
\affiliation{ICREA—Instituci{\'o} Catalana de Recerca i Estudis Avan{\c c}ats, 08015 Barcelona, Spain}

\author{Mohammad Hafezi}
\affiliation{Joint Quantum Institute, University of Maryland, College Park, MD 20742, USA}
\affiliation{Joint Center for Quantum Information and Computer Science, University of Maryland, College Park, MD 20742, USA}

\author{Andrey Grankin}
\affiliation{Joint Quantum Institute, University of Maryland, College Park, MD 20742, USA}

\date{\today}

\begin{abstract}

We propose that excitons in moir\'e transition metal dichalcogenide bilayers offer a promising platform for investigating collective radiative properties. While some of these optical properties resemble those of cold atom arrays, moir\'e excitons extend to the deep subwavelength limit, beyond the reach of current optical lattice experiments. Remarkably, we show that the collective optical properties can be exploited to probe certain correlated electron states without requiring subwavelength spatial resolution. Specifically, we illustrate that the Wigner crystal states of electrons doped into these bilayers act as an emergent periodic potential for excitons. Moreover, the collective dissipative excitonic bands and their associated Berry curvature can reveal various charge orders that emerge at the corresponding electronic doping. Our study provides a promising pathway for future research on the interplay between collective effects and strong correlations involving moir\'e excitons.

\end{abstract}

\maketitle
\end{CJK}

\textit{Introduction.} --- The collective behavior of optical emitters in two-dimensional lattices driven by dipole-dipole interaction has generated considerable interest. 
Emitters in such arrays display a range of intriguing properties, including perfect reflection and transmission, dressed lineshift, enhanced and suppressed radiation, and even topological attributes such as collective Chern bands and protected edge states~\cite{shahmoon2017cooperative,perczel2017photonic,perczel2017topological,moreno2021quantum}. Recently, this system has been experimentally realized using cold atoms in optical lattices~\cite{rui2020subradiant}, providing a promising avenue for the study of these coherent phenomena.

On the other hand, excitons (electron-hole bound states) in transition metal dichalcogenide (TMD) bilayers~\cite{rivera2018interlayer,jiang2021interlayer,kennes2021moire,huang2022excitons,mak2022semiconductor,du2023moire} can form in a moir\'e lattice, providing an interesting platform to study collective optical properties.
Moreover, the bilayers can host Wigner crystal (WC) states of doped electrons ~\cite{regan2020mott,tang2020simulation,xu2020correlated,jin2021stripe,miao2021strong,campbell2022exciton,xiong2023correlated,arsenault2024two,gao2024excitonic}, which can modify the excitonic properties.

\begin{figure}[t]
\centering
\includegraphics[width=\columnwidth]{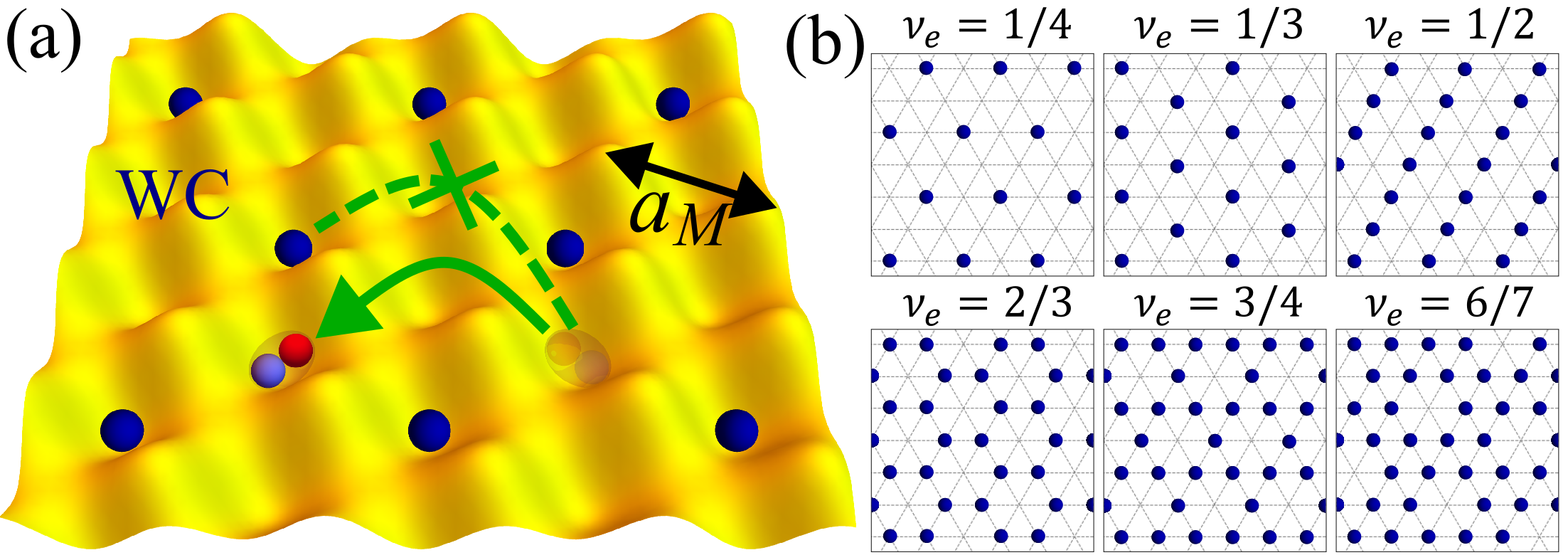}
\caption{
(a) Illustration of tunneling mediated by dipole-dipole interactions (green arrow) of an exciton (indicated by a pair of red and light blue dots) in a moir\'e potential (yellow) with period $a_M$. 
The potential can host a Wigner crystal state (WC) of doped electrons (dark blue dots), here a filling of $\nu_e=\frac{1}{3}$ is illustrated.
Due to exciton-electron repulsion, the exciton cannot populate the sites occupied by electrons (indicated by the crossed-out dashed path). The remaining available sites form an emergent excitonic lattice.
(b) WC at various $\nu_e$ in zero-twist WSe\textsubscript{2}/WS\textsubscript{2}~\cite{xu2020correlated,huang2021correlated}. 
The remaining empty sites provide a lattice for exciton tunneling shown in panel (a).
}
\label{Fig_1}
\end{figure}

In this work, we investigate the collective optical properties of moir\'e excitons in TMDs at incompressible states of doped electrons for various fractional fillings $\nu_e$~\footnote{For clarity, the term ``collective'' in this work refers to the interference effects of states of an optical emitter from different sites~\cite{scully2009collective,rohlsberger2010collective} rather than many-body physics of multiple emitters~\cite{gross1982superradiance,noh2016quantum,park2023dipole}}\nocite{scully2009collective,rohlsberger2010collective,gross1982superradiance,noh2016quantum,park2023dipole}. 
We focus on a weak excitation regime that yields dilute excitons, allowing us to neglect their electrostatic repulsion that manifests in high-power experiments~\cite{park2023dipole,gao2024excitonic,miao2021strong,xiong2023correlated}.
These excitations tunnel between empty moir\'e sites through dipole-dipole interaction, but cannot sit on top of a doped electron due to exciton-electron repulsion, as illustrated in Fig.~\ref{Fig_1}(a). 
Therefore, WCs emerging at finite $\nu_e$ act as additional superpotentials, modifying the lattice potential experienced by a single exciton.
Moreover, different $\nu_e$'s give rise to distinct WCs [see Fig.~\ref{Fig_1}(b)], which induce distinctive excitonic dynamics with quantitatively different collective (spectral and topological) excitonic properties~\cite{buckley2022optimized}. As such, those optical signatures serve as fingerprints of the corresponding WCs (see Table~\ref{tab:Wigner_crystal_WSe2WS2}).
Notably, the filling fractions $\nu_e$ can be easily accessed in gate-tunable experiments.
Such tunability, combined with adjustable twist angles, further allows for the exploration of subwavelength emitter-array physics within different lattice configurations.

We find strong cooperative effects for all emergent lattice structures considered in this work as these arrays are deeply subwavelength.
More specifically, the radiative decay rate of excitons within the light cone (LC) can experience a substantial enhancement that scales with the number of supersites within the resonant wavelength of the exciton $\lambda_{\mathrm{ex}}$~\cite{perczel2017photonic,perczel2017topological}, which, to the best of our knowledge, has not been considered in previous analyses of moiré excitons.
This increase could possibly bridge the discrepancy between the experimental values of exciton linewidths ($\sim 1$meV~\cite{tang2020simulation,xu2020correlated}) and their estimations via Wigner-Weisskopf theory~\cite{steck2007quantum} ($\sim10^{-6}-10^{-3}$meV~\footnote{In this estimation, we pick the dielectric constant to be $1-10$ and the transition dipole to be $0.1-1$ e$\cdot$nm, which are standard for TMD bilayers~\cite{park2023dipole,barre2022optical}}). 
In addition, we find that (the overall trend of) the radiative linewidth of the lowest energy exciton within charge-ordered states decreases with $\nu_e$, which serves as a signature for different Wigner crystal (WC) states. Interestingly, we further find the emergence of nontrivial Berry curvature for certain lattice structures in the absence of time-reversal symmetry. The dependence of Berry curvature on the emergent lattice at different $\nu_e$ offers a potential new probe for experimentally characterizing various WCs (see Table~\ref{tab:Wigner_crystal_WSe2WS2}).
Finally, we demonstrate that both the collective radiative linewidth and the Berry curvature can be extracted from polarization- and momentum-resolved far-field reflection measurements~\cite{gianfrate2020measurement}, which could provide more information compared to previous optical experiments~\cite{xu2020correlated,huang2021correlated,smolenski2021signatures,zhou2021bilayer}.

\begin{table}[t]
\centering
\begin{tabular}{|c|c|c|c|}
\hline
Emergent arrays
&
$\nu_e$
&
Superradiant $\Lambda$
&
$C_{\mathrm{LC}}$
\\
\hline
Triangular
&
$0,\frac{2}{3},\frac{3}{4},\frac{6}{7}$
&
$0,1$
&
$-1,1$
\\
\hline
Rectangular 
&
$\frac{1}{2}$
&
$0,1$
&
$0,0$
\\
\hline
Honeycomb 
&
$\frac{1}{3}$
&
$0,1$
&
$-1,1$
\\
\hline
Kagome
&
$\frac{1}{4}$
&
$0,1,3,4$
&
$-1,1,1,-1$
\\
\hline
\end{tabular}
\caption{Emergent lattices and collective emission properties of interlayer excitons at various filling fractions ($\nu_e$), which form complementary lattices of charge orders in zero-twist WSe\textsubscript{2}/WS\textsubscript{2}~\cite{xu2020correlated,huang2021correlated}. 
The third column lists the collective band indices (in energy order) $\Lambda$ with enhanced radiative decay (within the light cone) compared to the bare dipole transition rate $\gamma$. 
$C_{\mathrm{LC}}$ denotes the corresponding Berry curvature, summed over the light cone, upon an out-of-plane magnetic field with Zeeman splitting $\mu_BB=20\gamma$.
}
\label{tab:Wigner_crystal_WSe2WS2}
\end{table}

\textit{Two-band model.} --- We consider the lowest conduction (CB) and the highest valence (VB) bands of two TMD monolayers.
Each band is labeled by the valley pseudospin, which is locked to the real spin degrees of freedom for energy scales lower than the corresponding spin-orbit splittings~\cite{xiao2012coupled}.
Stacking these layers with a tunable twist angle or (for heterobilayers) lattice mismatch generally leads to interlayer coupling with an enlarged spatial periodicity $a_M$.
We specifically consider heterobilayers~\cite{kang2013band}, as the band offsets therein allow us to capture the effect of interlayer tunneling by emergent superlattice potentials, which split the original electronic dispersion into moir\'e bands~\footnote{Such band misalignment also suppresses the topological effects favored by layer-degenerate states~\cite{wu2017topological}, which is not the main focus of our work.}. 
In addition, we consider doping electrons into the first conduction moir\'e band and focus on the regime with dilute electron-hole pairs~\cite{rivera2018interlayer,naik2022intralayer}, generated via weak optical excitation.
Note that within such a dilute regime, the emission properties are captured by the single excitation subspace, in which the spectrum of a master equation is mathematically equivalent to the eigenspectrum of a non-Hermitian Hamiltonian~\cite{shahmoon2017cooperative,perczel2017photonic,perczel2017topological}.
While our formalism can be straightforwardly generalized, for simplicity, all (doped and optically excited) electrons are assumed to be in the same layer.

We start with the following two-band model in the absence of an external drive (see its microscopic origin in Ref.~\cite{Supplement}\nocite{Haug2004,tran2019evidence,karni2022structure,abrikosov2012methods}):
\begin{equation}
\label{eq:H_two_band}
\hat{\mathcal{H}}
=
\sum_{n=c,v}
\sum_{\tau}
\int d^2\bm{s}
\hat{\psi}_{\tau}^{(n)\dagger}(\bm{s})
h_n(\bm{s})
\hat{\psi}_\tau^{(n)}(\bm{s})
+
\hat{\mathcal{V}}_e
+
\hat{\mathcal{V}}_d
.
\end{equation}
Here, $n$ labels the (monolayer) band index with $c$ and $v$ indicating CB and VB, respectively; $\tau\in\pm$ represents the valley pseudospin, and $\bm{s}$ is the in-plane continuous position variable.
$\hat{\psi}_\tau^{(c)}(\bm{s})$ and $\hat{\psi}_\tau^{(v)}(\bm{s})$ denote the corresponding annihilation operators for CB electrons and VB \textit{holes}, respectively.
$h_n(\bm{s})=\frac{\hat{p}_{n,\tau}^2}{2m_n}+\Delta_n(\bm{s})+(\delta_{n,c}-\delta_{n,v})(E_n^{0}+\mu)$ is the energy operator describing the non-interacting sector of the charge dynamics on top of the superlattice potential $\Delta_n(\bm{s})$, with $\hat{p}_{n,\tau}$, $m_n$, $E_n^0$, and $\mu$ being the momentum operator relative to the valleys, the effective mass, energy offset at the valley momentum, and chemical potential, respectively.
$\hat{\mathcal{V}}_e = \frac{e^2}{8\pi\epsilon} \int_{\bm{s},\bm{s}'}  \frac{:\hat{\rho}_{c}(\bm{s})\hat{\rho}_{c}(\bm{s}'):}{|\bm{s}-\bm{s}'|}-\frac{2\hat{\rho}_{c}(\bm{s})\hat{\rho}_{v}(\bm{s}')}{\sqrt{(\bm{s}-\bm{s}')^2+z_{cv}^2}}$ is the interaction between fermion densities $\hat{\rho}_n(\bm{s})=\sum_\tau\hat{\psi}_{\tau}^{(n)\dagger}(\bm{s})\hat{\psi}_{\tau}^{(n)}(\bm{s})$~\footnote{Similar interactions between the dilute valence holes are neglected.} with $\int_{\bm{s},\bm{s}'} =\int d^2\bm{s}d^2\bm{s}'$ and the colons indicating normal-ordering of the operators in between.
$-e$, $\epsilon$, and $z_{cv}$ are the electron charge, static electric permittivity, and the out-of-plane distance between CB electrons and VB holes~\footnote{$z_{cv}=0$ if they are in the same layer}, respectively.
In addition to $\hat{\mathcal{V}}_e$, electron-hole pairs can also interact through their transition dipoles, yielding:
\begin{equation}
\label{eq:V_d}
\hat{\mathcal{V}}_d
=
\frac{\omega_p^2}{c^2\epsilon_0}
\int_{\bm{s},\bm{s}'}
\hat{\bm{P}}^\dagger(\bm{s})
\cdot
\bm{\mathcal{G}}\left(\omega_p,\bm{s}-\bm{s}'\right)
\cdot
\hat{\bm{P}}(\bm{s}')
,
\end{equation}
where $c$ is the speed of light and $\epsilon_0$ is the vacuum electric permittivity~\footnote{Dielectric screening is neglected at the target frequency for simplicity.}.
$\hat{\bm{P}}^\dagger(\bm{s})
=\sum_{\tau}\bm{d}_{\tau}^{cv}\hat{\psi}_{\tau}^{(c)\dagger}(\bm{s})\hat{\psi}_{\tau}^{(v)\dagger}(\bm{s})$ is the pair creation operator~\footnote{R-stacked bilayers are considered such that the two charges have the same $\tau$~\cite{seyler2019signatures}.}, where $\bm{d}_{\tau}^{cv}$ is the transition dipole matrix element between the CB and VB at valley $\tau$.
$\bm{\mathcal{G}}\left(\omega_p,\bm{s}-\bm{s}'\right)$ is the dyadic Green's tensor evaluated at the frequency of the target pair state $\omega_p$~\cite{Supplement}.
In the following, we use the two-band model $\hat{\mathcal{H}}$ to construct the superlattice Hamiltonian for optical excitations in the lowest energy manifold and doped electrons.

\textit{Zero doping.} --- At $\nu_e=0$, optical excitations from Eq.~\eqref{eq:H_two_band} are given by excitons.
To capture the lowest composite particle, we employ the tight-binding approximation to the two-particle Hamiltonian operator from $\hat{\mathcal{H}}$~\footnote{In the tight-binding regime, both the center-of-mass spatial fluctuations of the electron-hole pair from the moiré potential minima and the electron-hole separation are much smaller than the moiré period.}, whose eigenfunctions in the tight-binding limit can be approximated by moir\'e-Wannier orbitals~\cite{Supplement}.
The lowest energy orbital at each supersite $\bm{R}$, denoted as $w_{\bm{R}}(\bm{s}_c,\bm{s}_v)$ with $\bm{s}_c$ and $\bm{s}_v$ being the coordinates of the electron and hole, respectively, defines the corresponding exciton creation operator:
\begin{equation}
\label{eq:x_R_tau}
\hat{x}_{\bm{R},\tau}^\dagger
=
\int
d^2\bm{s}_c
d^2\bm{s}_v
w_{\bm{R}}(\bm{s}_c,\bm{s}_v)
\hat{\psi}_\tau^{(c)\dagger}(\bm{s}_c)
\hat{\psi}_\tau^{(v)\dagger}(\bm{s}_v)
.
\end{equation}
Projecting $\hat{\mathcal{H}}$ onto these single-exciton basis states, the superlattice Hamiltonian becomes~\footnote{Here, we drop the dipole-dipole interaction from the small out-of-plane \textit{transition} dipole (not the permanent dipole) of interlayer excitons~\cite{yu2015anomalous}, which can be further suppressed via hybridization with intralayer excitons~\cite{alexeev2019resonantly}.}:
\begin{equation}
\label{eq:H_inplane_undoped}
\hat{H}
=
\left(
\omega_{\mathrm{ex}}
-
\frac{i\gamma}{2}
\right)
\hat{n}_{\mathrm{ex}}
-
\sum_{\bm{R}\neq\bm{R}'}
\sum_{\tau,\tau'}
t_{\bm{R},\bm{R}'}^{\tau,\tau'}
\hat{x}_{\bm{R},\tau}^\dagger
\hat{x}_{\bm{R}',\tau'}
,
\end{equation}
where $\hat{n}_{\mathrm{ex}}=\sum_{\bm{R},\tau}\hat{x}_{\bm{R},\tau}^\dagger\hat{x}_{\bm{R},\tau}$.
$\omega_{\mathrm{ex}}$ and $\gamma$ denote the exciton frequency and decay rate (we set $\hbar=1$ hereafter), respectively, in a unit supercell problem, whereas the true excitation spectrum in the superlattice is renormalized by the tunneling:
\begin{equation}
\label{eq:tunneling}
t_{\bm{R},\bm{R}'}^{\tau,\tau'}
=
t_{\bm{R},\bm{R}'}^{\mathrm{int}}
\delta_{\tau,\tau'}
-
\frac{\omega_{\mathrm{ex}}^2|d|^2}{c^2\epsilon_0}
\bm{e}_{\tau}^\ast \cdot
\bm{\mathcal{G}}^{\mathrm{ex}}\left(\bm{R}-\bm{R}'\right)
\cdot \bm{e}_{\tau'}
,
\end{equation}
which incorporates the sector from charge dynamics (denoted as $t_{\bm{R},\bm{R}'}^{\mathrm{int}}$) and the dipole-dipole interaction from the in-plane components of $\bm{d}_\tau^{cv}$~\footnote{Similar interactions between dipoles within the same supercell has already been incorporated in $\omega_{\mathrm{ex}}$ and $\gamma$.}, denoted as $d\bm{e}_\tau^\ast$ with $\bm{e}_\tau=\frac{\bm{e}_x+i\tau\bm{e}_y}{\sqrt{2}}$ ($\bm{e}_x$ and $\bm{e}_y$ are in-plane unit vectors)~\cite{wu2018theory,yu2015anomalous}.
Here, $\bm{\mathcal{G}}^{\mathrm{ex}}\left(\bm{R}-\bm{R}'\right)=\int_{\bm{s},\bm{s}'}w_{\bm{R}}^{\ast}(\bm{s})\bm{\mathcal{G}}\left(\omega_{\mathrm{ex}};\bm{s}-\bm{s}'\right)w_{\bm{R}'}(\bm{s}')$ where $w_{\bm{R}}(\bm{s})\equiv w_{\bm{R}}(\bm{s},\bm{s})$ is approximated as a Gaussian with width $a_W$ for simplicity.
We refer to the supplementary material for details of $\hat{H}$~\cite{Supplement}.

\textit{Finite doping with charge order.} --- We can generalize Eq.~\eqref{eq:H_inplane_undoped} to finite $\nu_e$, where doped electrons form WCs~\footnote{The collective excitonic properties for hole-doped cases are qualitatively similar to electron-doped ones.}.
These charge orders could emerge from a generalized Hubbard model~\cite{pan2020quantum,tan2023doping}, which naturally appears by projecting Eq.~\eqref{eq:H_two_band} onto the first moir\'e band of doped charges~\cite{Supplement}.
Importantly, since the exciton satisfies the tight-binding criterion, its hole remains bound to its original electron, preventing recombination with doped charges~\cite{upadhyay2024giant}.
Further assuming that the electronic states being stable against the dynamics of dilute optical excitations, we can treat the WC as a spatially periodic detuning, consistent with experimental signatures~\cite{smolenski2021signatures,zhou2021bilayer}.
This detuning is characterized by the gap between $\omega_{\mathrm{ex}}$ and the energy of lowest three-body state (one doped electron together with the excited electron-hole pair) within a supercell, denoted as $\omega_t$.  
In particular, in the regime $|\omega_t-\omega_{\mathrm{ex}}|\gg|t_{\bm{R},\bm{R}'}^{\tau,\tau'}|$, the low-energy optical excitations are either within the $\sim\omega_{\mathrm{ex}}$ or $\sim\omega_t$ manifold, depending on the sign of $\omega_t-\omega_{\mathrm{ex}}$.
We focus specifically on the case $\omega_{\mathrm{ex}}\ll\omega_t$, which is realizable by interlayer excitons~\cite{Supplement,gao2024excitonic}, to study the collective behavior of these emitters (three-body states with $\omega_{\mathrm{ex}}\gg\omega_t$ correspond to trions~\cite{ciorciaro2023kinetic}, which is not the focus of this work).
In this subspace, these excitons are still described by the Hamiltonian Eq.~\eqref{eq:H_inplane_undoped}~\footnote{Note that the energy integrals are different for the excitations in the $\sim\omega_{\mathrm{ex}}$ and $\sim\omega_t$ manifolds, due to their distinction in wavefunctions}, except that now $\{\bm{R}\}$ lies in the emergent lattices, which are complementary lattices of the WCs.
Accordingly, different emergent lattices can be realized simply by accessing distinct WCs in moir\'e TMD bilayers, which can be tuned by the electron filling fraction $\nu_e$ (doped electrons per supercell)~\cite{pan2020quantum,xu2020correlated,huang2021correlated}.
We list a few possible emergent lattices observed in  WSe\textsubscript{2}/WS\textsubscript{2} in Table~\ref{tab:Wigner_crystal_WSe2WS2}.

\begin{figure}[t]
    \centering
    \includegraphics[width=\columnwidth]{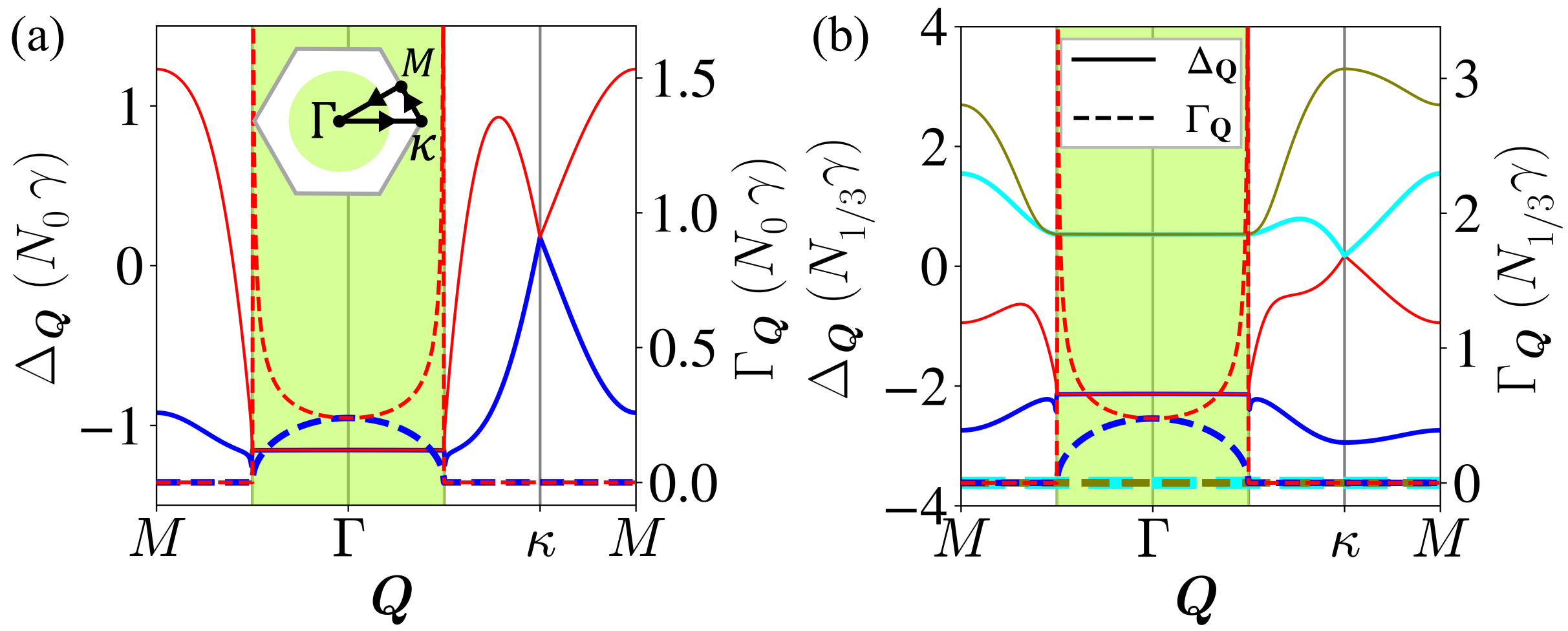}
    \caption{Collective excitonic lineshifts $\Delta_{\bm{Q}}$ (solid lines) and linewidths $\Gamma_{\bm{Q}}$ (dashed) emerging from charge-ordered zero-twist WSe\textsubscript{2}/WS\textsubscript{2} with electron fillings (a) $\nu_e=0$ and (b) $\nu_e=\frac{1}{3}$. 
    The vertical axes are displayed in units of $\gamma N_{\nu_e}$, with $N_{\nu_e}$ defined in Eq.~\eqref{eq:N_nu_e_theta}.
    The horizontal axes represent the Bloch momentum $\bm{Q}$, which follows a piecewise-linear path through high-symmetry points in the Brillouin zone, as indicated by the hexagon in the inset of (a).
    Momenta within the light cone are indicated by the green shaded area (size enlarged for clarity).
    Different colors label distinct single-particle exciton bands.
    The parameters used are: $a_M=8.25$nm, $\omega_{\mathrm{ex}} =1.55$eV, and $a_W=2$nm~\cite{Supplement,park2023dipole}.
    }
    \label{Fig_2}
\end{figure}

\textit{Collective bands.} --- We proceed to study the eigenspectrum of Eq.~\eqref{eq:H_inplane_undoped}.
To capture the relevant physics, we set $t_{\bm{R},\bm{R}'}^{\mathrm{int}}=0$ in Eq.~\eqref{eq:tunneling} throughout this work~\footnote{Neglecting $t_{\bm{R},\bm{R}'}^{\mathrm{int}}$ is a valid approximation at large $a_M$ because it decays exponentially with $|\bm{R}-\bm{R}|/a_M$, whereas the dipole-dipole interaction only scales as a power law.}.
Diagonalization of the Hamiltonian Eq.~\eqref{eq:H_inplane_undoped} yields the emitter spectrum $\omega_{\mathrm{ex}}+\Delta_{\bm{Q}}-\frac{i}{2}\Gamma_{\bm{Q}}$ characterized by the center-of-mass Bloch momentum $\bm{Q}$.
Note that, if the excitonic states at all $\bm{R}$ constructively interfere, $\Gamma_{\bm{Q}}$ should scale with $\gamma N_{\nu_e}(\theta)$~\cite{shahmoon2017cooperative}, where $N_{\nu_e}(\theta)$ is the ratio between the squared exciton wavelength and the unit cell area of the effective excitonic lattice at twisting angle $\theta$:
\begin{equation}
\label{eq:N_nu_e_theta}
N_{\nu_e}(\theta)
=
(1+\theta^2/\delta^2)
N_{\nu_e}
,\quad
N_{\nu_e}
=
\lambda_{\mathrm{ex}}^2/\mathcal{A}_{\nu_e}
.
\end{equation}
Here $\mathcal{A}_{\nu_e}$ denotes the emergent unit cell area at electron filling $\nu_e$ and zero twist and $\delta$ defines the lattice mismatch between the two monolayers~\cite{park2023dipole}.
We present the collective spectrum in units of $\gamma N_{\nu_e}(\theta)$ hereafter to indicate the extent of constructive interference of the exciton eigenstates.

In Fig.~\ref{Fig_2}, we illustrate the collective lineshifts $\Delta_{\bm{Q}}$ and linewidths $\Gamma_{\bm{Q}}$ for the untwisted WSe\textsubscript{2}/WS\textsubscript{2} bilayer at $\nu_e=0$ and $\frac{1}{3}$, corresponding to triangular and honeycomb lattices, respectively. 
The cases with Kagome and rectangular lattices can be found in the supplementary materials~\cite{Supplement}.
For all $\nu_e$ of interest, both quantities vary on scales of $N_{\nu_e}\gamma$ with $N_{\nu_e}\sim10^4$.
Qualitatively, this is because emissions from supersites much closer than the exciton wavelength $\lambda_{\mathrm{ex}}$ can constructively interfere.

The collective behavior is qualitatively distinct within and outside of the LC.
For all $\nu_e$, $\Gamma_{\bm{Q}}$ is significantly greater than $\gamma$ within the LC but is suppressed outside, consistent with the fact that only emitters in the LC couple to light and radiate due to momentum conservation.
Note also that such a large decay rate in the LC indicates that an emitter tends to radiate before it hops to other supersites, which is also reflected by the relatively flat $\Delta_{\bm{Q}}$ (with respect to $\bm{Q}$) therein.

\begin{figure}[t]
    \centering
    \includegraphics[width=\columnwidth]{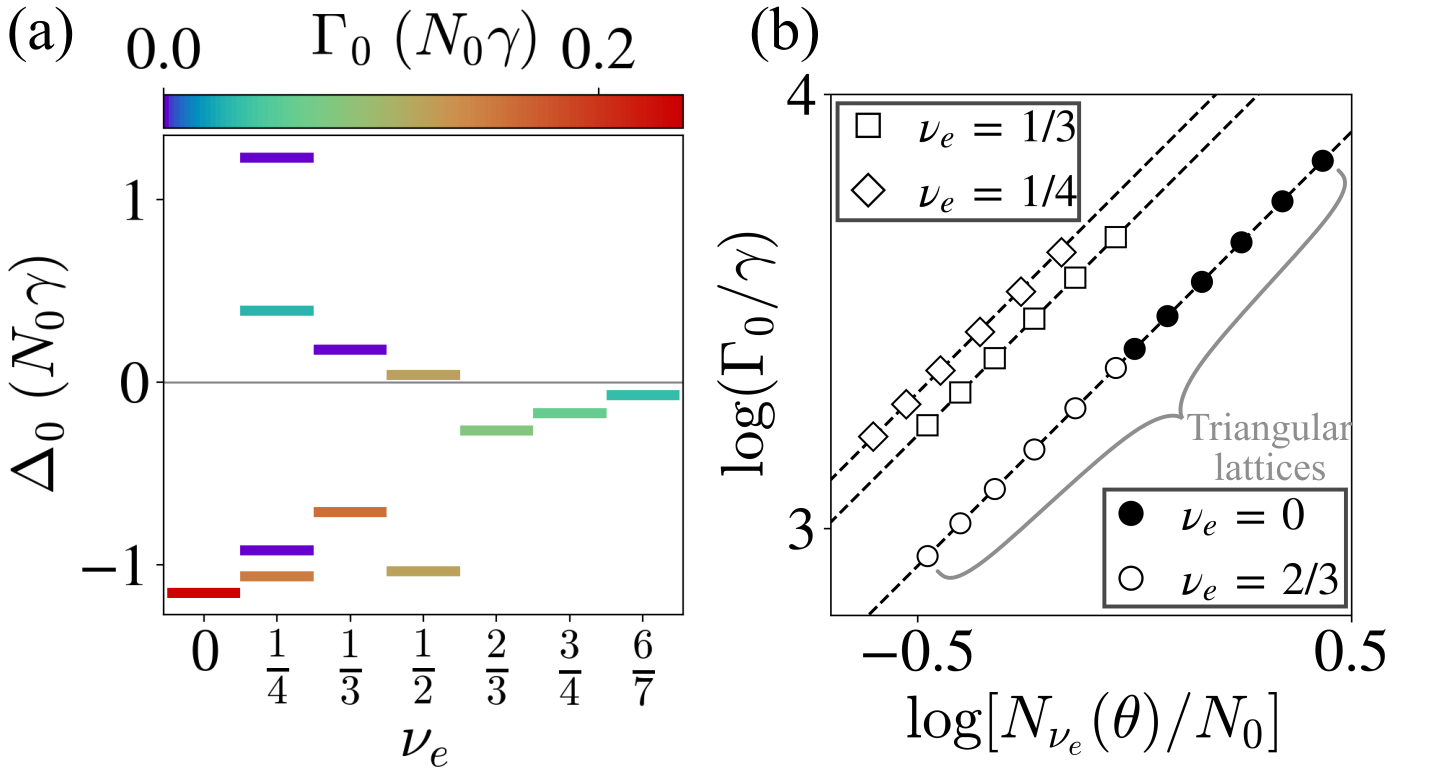}
    \caption{Dependence of collective lineshifts and linewidths at $\bm{Q}=0$ (a) on $\nu_e$ for all states at zero twist and (b) on twisting angle $\theta$ and $\nu_e$, whose effects manifest in the combined factor $N_{\nu_e}(\theta)$, see Eq.~\eqref{eq:N_nu_e_theta}.
    Dashed lines in (b) indicate linear fits with slope $1$.
    Here we set $\delta=0.04$~\cite{park2023dipole}, and $a_W(\theta)= (1+ \theta^2/\delta^2)^{-\frac{1}{4}}a_W(0)$~\cite{wu2018hubbard}; all other parameters are the same as in Fig.~\ref{Fig_2}.}
    \label{Fig_3}
\end{figure}

Aside from these common properties, collective bands exhibit several distinct features at different $\nu_e$.
We specifically illustrate how $\Delta_{0}$ and $\Gamma_{0}$ vary with $\nu_e$ in Fig.~\ref{Fig_3}(a), as these quantities are directly accessible via reflection measurements at normal incidence~\cite{shahmoon2017cooperative}.
First, different emergent lattices generally yield distinct numbers of bright levels, which serves as the most direct signature to identify certain WCs.
For instance, two \textit{bright} lines with equal and different $\Gamma_{0}$ can be associated to rectangular ($\nu_e = \frac{1}{2}$) and Kagome ($\nu_e = \frac{1}{4}$) lattices, respectively.
In contrast, triangular and honeycomb lattices cannot be differentiated by bright-level counting.

Another optical signature of the WCs is the linewidth $\Gamma_{0}$ of the lowest state at each $\nu_e$, which is typically the fastest radiative rate.
By comparing their values (which overall decreases with $\nu_e$), it is possible to determine whether the corresponding WCs are formed in experiments~\footnote{While a similar comparison can be made for $\Delta_{0}$, we anticipate that additional factors beyond the scope of this study, including long-range exciton-electron interactions, may also influence the exciton energy. In spectral measurements, isolating $\Delta_{0}$ from those factors could be challenging.}.

Finally, these spectral properties can also be tuned via twisting angle $\theta$.
Fig.~\ref{Fig_3}(b) illustrates the dependence of $\Gamma_0$ on $N_{\nu_e}(\theta)$.
Crucially, for the same type of lattice geometry, the collective linewidths scale inversely with the emergent unit cell area, which can be confirmed by the analytic expression of the Green's tensor at $\bm{Q}=0$.
These scalings provide an indirect probe for the emergent lattice geometry: for instance, all triangular arrays (e.g., $\nu_e=0,\frac{2}{3}$) belong to the same line in Fig.~\ref{Fig_3}(b).
Therefore, the intercept in this logarithmic plot serves as another fingerprint of the WCs.

\textit{Berry curvature.} --- Topological features can also emerge in these collective excitonic bands. 
To see this, we introduce a magnetic field to break time-reversal symmetry and compute the total Berry curvature within the LC, $C_{\mathrm{LC}}$, for (directly) optically accessible states, as summarized in Table~\ref{tab:Wigner_crystal_WSe2WS2}.
Excitonic bands arising from triangular, honeycomb, and Kagome effective lattices yield $C_{\mathrm{LC}} = \pm1$, whereas those from rectangular lattices remain topologically trivial.
This behavior originates from the phase winding of the light-mediated coupling between valley doublets, which effectively manifests only in lattices with $C_3$ rotational symmetry, rather than from topological electronic bands~\cite{xie2024long}.

Finally, we propose a far-field reflection experiment to detect $C_{\mathrm{LC}}$, where the scattering matrix element extracted from cross-circularly polarized input and output fields, $S_{-+}$, is measured.
Specifically, its momentum dependence, controlled by incidence direction, possesses difference number of nodes for excitonic bands with different topological features.
We leave the details to the End Matter.

\textit{Outlook.} --- 
Our formalism can be further generalized to describe the collective behavior of moir\'e excitons involving other strong correlations to study their interplay~\cite{bloch2022strongly}.
For instance, going beyond the single excitation subspace~\cite{pedersen2024green}, nonlinearities inherent in these emitters can lead to a variety of interesting physical phenomena, including Dicke superradiance~\cite{masson2022universality,sierra2022dicke,kumlin2024superradiance}, optical bistability~\cite{camacho2022moire}, leaky condensation~\cite{remez2022leaky}, and phase space filling~\cite{huang2023non,song2024microscopic,song2024electrically}.
Another intriguing problem is whether the spin correlations that emerge near half-filling $\nu_e=1$, such as the magnetic polaron effect~\cite{huang2023spin,huang2023mott} and kinetic magnetism~\cite{ciorciaro2023kinetic}, could play a role in the cooperative excitonic properties.

In addition to WCs where the doped electrons only act as an effective lattice potential to the excitons, it is natural to ask about their collective behavior at general $\nu_e$.
At fillings slightly away from the ones providing WCs, metastable frozen charge configurations could emerge and play the role of a random potential to emitters that breaks the translation symmetry~\cite{tan2023doping}.
We anticipate this randomness to suppress the degree of constructive interference, indicating weaker radiative decay.

Another possible outlook for moir\'e excitons is to simulate topological physics in two-dimensional dipolar spin systems~\cite{huang2023non}.
In particular, their tunneling driven by dipole-dipole interaction could provide relatively flat collective bands with nontrivial Berry curvature within the light cone.
These ingredients allow for the emergence of topological phases such as fractional Chern insulators and spin liquids~\cite{yao2013realizing,yao2018quantum,perczel2020topological}.
We therefore anticipate these optical excitations in TMD bilayers to act as a platform for these phases of matter.

\textit{Acknowledgements.} --- 
We acknowledge A. Srivastava, S. Yelin and A. Asenjo-Garcia and D. Goncalves for useful discussions. The work at Maryland was supported by W911NF2010232, MURI FA9550-19-1-0399, DARPA HR00112530313 and Simons and Minta Martin Foundations. Y.-X.W.~acknowledges support from a QuICS Hartree Postdoctoral Fellowship. Y.-Q. W. acknowledges the support from the JQI postdoctoral fellowship at the University of Maryland.
D.E.C. acknowledges support from the European Union, under European Research Council grant agreement No 101002107 (NEWSPIN), FET-Open grant agreement No 899275 (DAALI) and EIC Pathfinder Grant No 101115420 (PANDA); the Government of Spain (Severo Ochoa Grant CEX2019-000910-S [MCIN/AEI/10.13039/501100011033]); QuantERA II project QuSiED, co-funded by the European Union Horizon 2020 research and innovation programme (No 101017733) and the Government of Spain (European Union NextGenerationEU/PRTR PCI2022-132945 funded by MCIN/AEI/10.13039/501100011033); Generalitat de Catalunya (CERCA program and AGAUR Project No. 2021 SGR 01442); Fundació Cellex, and Fundació Mir-Puig. All authors acknowledge the hospitality of the Kavli Institute for Theoretical Physics (KITP) supported by grant NSF PHY-1748958.

\bibliography{Biblio}

\appendix

\clearpage

\onecolumngrid

\section{End Matter}

\twocolumngrid

\begin{figure}[t]
    \centering
    \includegraphics[width=\columnwidth]{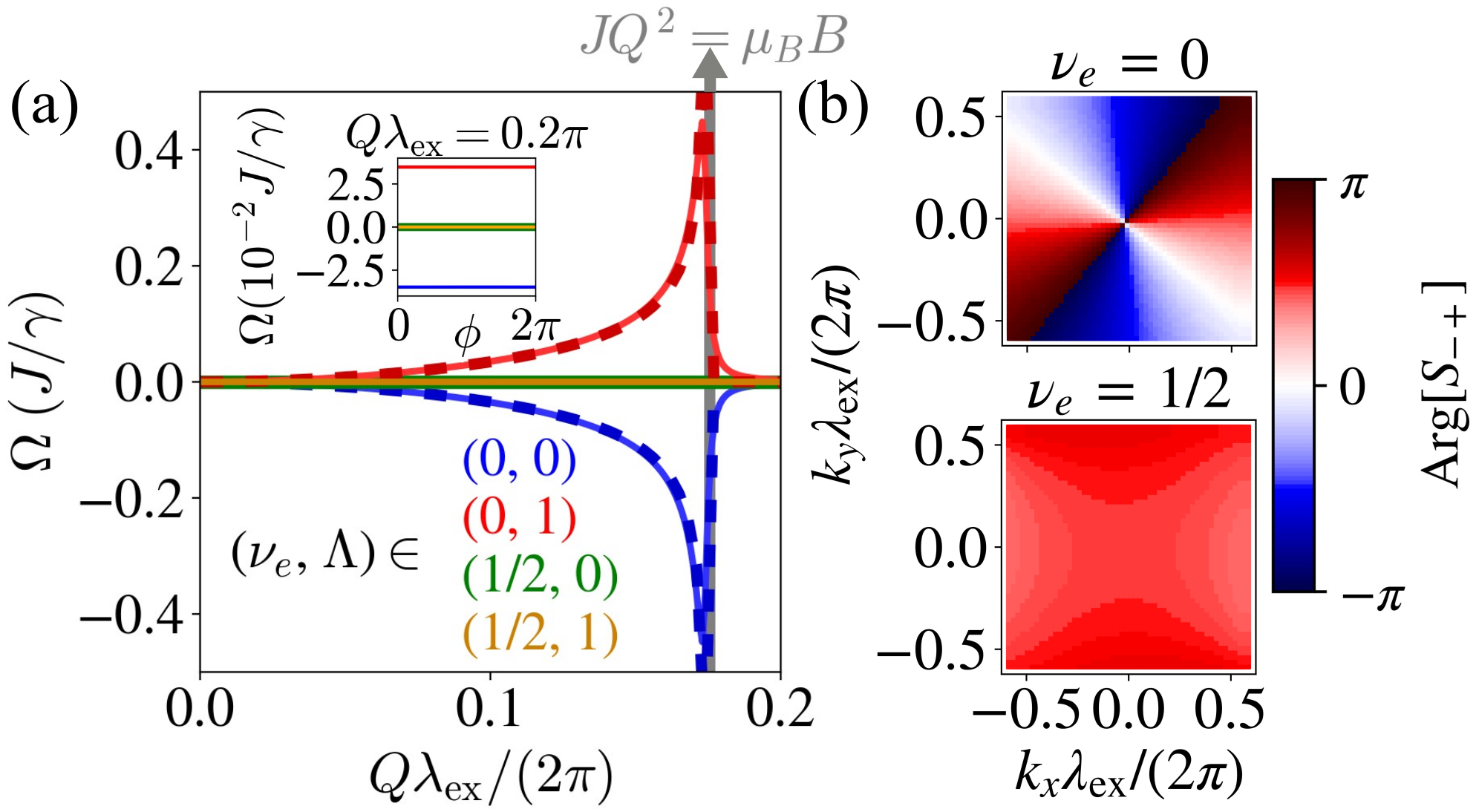}
    \caption{Comparison between properties of topological and non-topological collective bands from $\nu_e=0$ and $\nu_e=\frac{1}{2}$, respectively, with Zeeman splitting $\mu_BB=20\gamma$.
    (a) Dependence of Berry curvature ($\Omega$) on Bloch momentum magnitude $Q$ for the two bands at each $\nu_e$ (labeled by colors and $\Lambda$), with solid and dashed lines indicating results from the full model $\hat{H}'$ and the low-momentum model Eq.~\eqref{eq:low_kB_model}, respectively.
    The vertical axis is in units of $J/\gamma$, where $J$ is fit from $\Gamma_{\bm{Q}}$ in Fig.~\ref{Fig_2}(a) using a momentum sample within $Q\leq {\pi}/{25\lambda_{\mathrm{ex}}}$.
    The gray vertical line indicates the momentum satisfying $JQ^2=\mu_BB$.
    The inset shows that the dependence of $\Omega$ on the polar angle of $\bm{Q}$ is negligible. 
    (b) The phase of $S_{-+}$ at $k=\frac{2\pi}{\lambda_{\mathrm{ex}}}$ and $\bm{k}_{||}=k_x\bm{e}_x+k_y\bm{e}_y$ for each $\nu_e$.
    All other parameters are the same as in Fig.~\ref{Fig_2}.}
    \label{Fig_4}
\end{figure}

\textit{Appendix A: Details of Berry curvature.} ---
Incorporating an out-of-plane external magnetic field $B$ adds a Zeeman splitting between the valley-degenerate doublets to Eq.~\eqref{eq:H_inplane_undoped}.
We accordingly proceed with the Hamiltonian $\hat{H}\to\hat{H}'\equiv\hat{H}+\mu_BB\sum_{\tau = \pm,\bm{R}}\tau\hat{x}_{\bm{R},\tau}^\dagger\hat{x}_{\bm{R},\tau}$, where $\mu_B$ denotes the magnetic dipole.
The Berry curvature $\Omega({\bm Q})=i\nabla_{\bm{Q}}\times\langle\bm{Q}|\nabla_{\bm{Q}}|\bm{Q}\rangle$ is computed for each band at several $\nu_e$, where $|\bm{Q}\rangle$ are right eigenstates of $\hat{H}'$.
Note that without loss of generality, the right eigenvectors of the non-Hermitian Hamiltonian are chosen to define the Berry curvature~\cite{shen2018topological,zhang2019non} (a more rigorous study of non-Hermitian topology is left for future work).
We specifically focus on $\Omega({\bm Q})$ of (directly) optically accessible states (i.e., within the LC) and compute the numerical results for $\nu_e=0$ and $\frac{1}{2}$ [see Fig.~\ref{Fig_4}(a)], evaluated under a standard momentum discretization scheme~\cite{fukui2005chern}.

At $\nu_e=0$, both collective excitonic states (labeled with respect to their energies by $\Lambda=0,1$) exhibit nontrivial and opposite $\Omega({\bm Q})$ centered at $\bm{Q}=0$.
Notably, the summation of $\Omega({\bm Q})$ in LC, $C_{\mathrm{LC}}$, for these bands takes integer values $\mp1$.
As briefly mentioned in the main text, topological features here originates from the phase winding of $\bm{e}_{\tau}^\ast \cdot
\bm{\mathcal{G}}^{\mathrm{ex}}\left(\bm{R}-\bm{R}'\right)
\cdot \bm{e}_{-\tau}$, which appears as the off-diagonal terms in the effective low momentum approximation of $\hat{H}'$, denoted as $\hat{h}(\bm{Q})$~\cite{Supplement}.
With $\bm{Q}=Q(\cos\phi\bm{e}_x+\sin\phi\bm{e}_y)$, where $\phi$ is the polar angle of $\bm{Q}$, we have:
\begin{equation}
\label{eq:low_kB_model}
\hat{h}(\bm{Q})
\simeq
\omega_{\mathrm{ex}} + \Delta_0 - \frac{i\Gamma_0}{2}
+
\begin{bmatrix}
\mu_BB & iJQ^{2} e^{-2i\phi}
\\
iJQ^{2} e^{2i\phi} & -\mu_BB
\end{bmatrix}
,
\end{equation}
where $J$ is a (generally complex) coefficient characterizing the band curvature around $\bm{Q}=0$.
We find that Eq.~\eqref{eq:low_kB_model} can approximately capture $\Omega(\bm{Q})$, particularly the sharp feature at $Q = \sqrt{|\mu_BB/J|}$.

In contrast, $C_{\mathrm{LC}}$ from generic arrays (see table~\ref{tab:Wigner_crystal_WSe2WS2}) may be trivial if the low-$Q$ model of the target band is different from Eq.~\eqref{eq:low_kB_model}.
For instance, both collective excitonic bands from the rectangular lattice at $\nu_e=\frac{1}{2}$ give zero Berry curvature, and hence $C_{\mathrm{LC}}=0$.
This is due to the absence of $C_{3}$ rotational symmetry, leading to nonzero off-diagonal terms in the corresponding Hamiltonian at $Q=0$, which indicates a gap separating two topologically trivial bands~\cite{Supplement}.

Similar arguments also apply to Berry curvatures of higher collective states.
More specifically, the suppressed $J$ for $\Lambda=2,3$ at $\nu_e=\frac{1}{3}$ and $\Lambda=2,5$ at $\nu_e=\frac{1}{4}$ (c.f.~the flat real and imaginary spectra in Fig.~\ref{Fig_2} and Supplementary Material~\cite{Supplement}) is consistent with the fact that $C_{\mathrm{LC}}=0$ in these bands (not shown).
In contrast, the $\Lambda=3,4$ states at $\nu_e=\frac{1}{4}$ exhibit a nonzero curvature in $\Gamma_{\bm{Q}}^{\Lambda}$ such that they sustain nonzero $\Omega$ in the LC.

\textit{Appendix B: Extracting band properties via optical reflectivity.} ---
In this appendix, we discuss optical experiments to probe the aforementioned collective excitonic properties.
The numerical results of Berry curvature for $\nu_e = 0$ and $\frac{1}{2}$ are shown in Fig.~\ref{Fig_4}(a), and their manifestation in optical measurements is illustrated in Fig.~\ref{Fig_4}(b).
Here, far-field reflection experiment with plane-wave incident light near exciton resonances is considered, and the target physical observables are the amplitude and phase of the scattered field, which can be obtained via standard homodyne technique~\cite{steck2007quantum}.
Below we illustrate that information about $\hat{H}'$ can be inferred from the scattering matrix extracted from such measurement.
For Bravais lattices (e.g., triangular and rectangular arrays), components of this matrix satisfy:
\begin{equation}
\label{eq:S_gD}
S_{\tau,\tau''}(\bm{k})
=
\frac{3\pi c\gamma}{\omega_{\mathrm{ex}}}
\sum_{\tau'}
g_{\tau,\tau'}(k,\bm{k}_{||})
D_{\tau',\tau''}(ck,\bm{k}_{||})
,
\end{equation}
where $\bm{k}_{||}$ is the in-plane component of $\bm{k}$ and $\tau$ ($\tau'$) stands for the (circular) polarization of light here (which is locked to the valley index of electrons in TMDs~\cite{rivera2018interlayer}).
$g_{\tau,\tau'}(k,\bm{Q}) = - \frac{i}{2\mathcal{A}_{\nu_e}} \frac{k^2\delta_{\tau,\tau'}-(\bm{e}_\tau^\ast\cdot\bm{Q})(\bm{e}_{\tau'}\cdot\bm{Q})}{k^2\sqrt{k^2-Q^2}}$ is the Fourier transform of the Green's tensor~\cite{Supplement}.
$D_{\tau,\tau'}(\omega,\bm{Q})=\langle\hat{x}_{\bm{Q},\tau}(\omega-\hat{H}')^{-1} \hat{x}_{\bm{Q},\tau'}^\dagger\rangle $ describes the collective response susceptibility of the exciton, evaluated in the vacuum state, with $\hat{x}_{\bm{Q},\tau}$ denoting the Fourier transform of $\hat{x}_{\bm{R},\tau}$. 
Once $\hat{H}'$ is reconstructed from $S_{\tau,\tau'}(\bm{k})$ via Eq.~\eqref{eq:S_gD}, the corresponding spectral and topological excitonic properties within the LC could be determined.
In particular, $C_{\mathrm{LC}}$ can be extracted from (half of) the winding number of phase of $S_{\tau,-\tau}$ with respect to the polar angle of $\bm{k}_{||}$, as shown in Fig.~\ref{Fig_4}(b).

For non-Bravais lattices, however, $S_{\tau,\tau'} (\bm{k})$ can only provide partial information about the exciton Hamiltonian because far-field measurements cannot resolve the sublattices. 
More specifically, $(\omega-\hat{H}')^{-1}$ in $D_{\tau,\tau'}(\omega,\bm{Q})$ has to be sublattice-symmetrized in this situation~\cite{Supplement}.
Therefore, measuring $S_{\tau,\tau'} (\bm{k})$ generally cannot uniquely determine all the excitonic properties within the LC.
One special case where such a scheme is still applicable is the honeycomb lattice, where sublattice symmetrization selects the $\Lambda=0,1$ doublet such that their $\Delta_{\bm{Q}}$, $\Gamma_{\bm{Q}}$, and $\Omega(\bm{Q})$ are still fully recoverable from reflection experiments~\footnote{Note that radiation from $\Lambda=2,3$ states of the honeycomb lattice is suppressed and hence its contribution to the optical response is negligible for all practical purposes}.
A complete reconstruction of the excitonic band structures would require additional measurements beyond $S_{\tau,\tau'}(\bm{k})$, which we leave for future work.

\end{document}


\begin{CJK}{UTF8}{gbsn}

\title{Supplementary Material: Collective optical properties of moire excitons}

\author{Tsung-Sheng Huang}
\affiliation{Joint Quantum Institute, University of Maryland, College Park, MD 20742, USA}

\author{Yu-Xin Wang (王语馨)}
\affiliation{Joint Center for Quantum Information and Computer Science, University of Maryland, College Park, MD 20742, USA}

\author{Yan-Qi Wang}
\affiliation{Joint Quantum Institute, University of Maryland, College Park, MD 20742, USA}

\author{Darrick Chang}
\affiliation{
ICFO-Institut de Ciencies Fotoniques, The Barcelona Institute of Science and Technology,
08860 Castelldefels (Barcelona), Spain}
\affiliation{ICREA—Instituci{\'o} Catalana de Recerca i Estudis Avan{\c c}ats, 08015 Barcelona, Spain}

\author{Mohammad Hafezi}
\affiliation{Joint Quantum Institute, University of Maryland, College Park, MD 20742, USA}
\affiliation{Joint Center for Quantum Information and Computer Science, University of Maryland, College Park, MD 20742, USA}

\author{Andrey Grankin}
\affiliation{Joint Quantum Institute, University of Maryland, College Park, MD 20742, USA}

\date{\today}

\maketitle
\end{CJK}

\section{Derivation of the two band model from microscopic Hamiltonian}

In this section, we present the derivation of the two band Hamiltonian starting from the following microscopic model within the Coulomb gauge (we set $\hbar = 1$ throughout the Supplementary Material for simplicity)~\cite{steck2007quantum}:
\begin{equation}
\label{eq:microscopic model}
\hat{\mathcal{H}}
=
\hat{\mathcal{H}}_0
+
\hat{\mathcal{H}}_L
+
\hat{\mathcal{H}}_T
,\;
\hat{\mathcal{H}}_0
=
\int
d^3\bm{r}
\hat{\psi}^\dagger(\bm{r})
\left[
-
\frac{\nabla_{\bm{r}}^2}{2m_0}
+
V_{\mathrm{at}}(\bm{r})
+
\mu
\right]
\hat{\psi}(\bm{r})
,\;
\hat{\mathcal{H}}_L
=
\frac{e^2}{2\epsilon_0}
\int d^3\bm{r}d^3\bm{r}'
\frac{
\hat{\psi}^\dagger(\bm{r})
\hat{\psi}^\dagger(\bm{r}')
\hat{\psi}(\bm{r}')
\hat{\psi}(\bm{r})
}{|\bm{r}-\bm{r}'|}
,
\end{equation}
where $m_0$, $-e$, and $\epsilon_0$ are the free electron mass, the electron charge, and the vacuum permittivity, respectively.
$\hat{\psi}(\bm{r})$ is the electron field operator at the \textit{three}-spatial-dimensional coordinate $\bm{r}$.
$V_{\mathrm{at}}(\bm{r})$ is the potential from the atoms of the two TMD layers, and $\mu$ is the chemical potential of electrons.
$\hat{\mathcal{H}}_T$ includes the Maxwell Hamiltonian and minimal light-matter coupling with respect to \textit{transverse} electromagnetic fields (see Section~\ref{sec:interaction}), whereas the longitudinal sector is packed into $\hat{\mathcal{H}}_L$ within the Coulomb gauge.

To further simplify this microscopic model, we decompose the electron field operator in terms of Bloch states states of two (decoupled) monolayers, which are respectively labeled by their orbitals, valley pseudospins (denoted with $\tau\in\pm$), spins, and Bloch momenta.
We aim at the low energy version of the microscopic model, where spin and valley indices are locked due to spin-orbit coupling in TMDs and only the lowest conduction and the highest valence orbitals of each layer are relevant (denoted as $c$ and $v$ respectively).
In addition, we assume that all (doped and optically generated) conduction electrons are within the same layer for simplicity (and similarly for valence holes) such that the layer label is locked to the band index.
With these considerations, we express:
\begin{equation}
\label{eq:electron_field}
\hat{\psi}(\bm{r})
\simeq
\sum_{\tau}
\sum_{\bm{p}_c}
\psi_{\tau\bm{K}_c+\bm{p}_c}^{c}(\bm{r})
\hat{c}_{\tau\bm{K}_c+\bm{p}_c}
+
\sum_{\bm{p}_v}
\psi_{\tau\bm{K}_v+\bm{p}_v}^{v}(\bm{r})
\hat{v}_{-\tau\bm{K}_v-\bm{p}_v}^\dagger
,
\end{equation}
where $\hat{c}_{\tau\bm{K}_c+\bm{p}_c}$ and $\hat{v}_{-\tau\bm{K}_v-\bm{p}_v}^\dagger$ are the conduction electron annihilation and valence \textit{hole} creation operators (hence momentum of the latter is flipped), with their corresponding electronic Bloch wavefunctions being $\psi_{\tau\bm{K}_c+\bm{p}_c}^{c}(\bm{r})$ and $\psi_{\tau\bm{K}_v+\bm{p}_v}^{v}(\bm{r})$.
We focus on the momenta near the Brillouin zone corners $\tau\bm{K}_n$ ($n\in\{c,v\}$) with $\bm{p}_n$ being the momentum relative to them.
At small twist and lattice mismatch such that the monolayer lattice vectors can be expressed as $\bar{\bm{a}}+\delta\bm{a}_n$ with $|\delta\bm{a}_n|\ll|\bar{\bm{a}}|$, and with the further assumption that the bilayer of interest is R-stacked, $\bm{K}_n\simeq \bar{\bm{K}} \equiv\frac{4\pi\bar{\bm{a}}}{3\bar{\bm{a}}^2}$.

\subsection{Non-interacting matter Hamiltonian}

We first utilize Eq.~\eqref{eq:electron_field} to simplify $\hat{\mathcal{H}}_0$.
By definition, the sector of each monolayer in $\hat{\mathcal{H}}_0$ is diagonalized by the Bloch states but leaves an interlayer coupling $\hat{V}_{IL}$, giving:
\begin{equation}
\hat{\mathcal{H}}_0
=
\sum_{\bm{p}_c,\tau}
E_{c}(\tau\bm{K}_c+\bm{p}_c)
\hat{c}_{\tau\bm{K}_c+\bm{p}_c}^\dagger
\hat{c}_{\tau\bm{K}_c+\bm{p}_c}
+
\sum_{\bm{p}_v,\tau}
E_{v}(\tau\bm{K}_v+\bm{p}_v)
\hat{v}_{-\tau\bm{K}_v-\bm{p}_v}
\hat{v}_{-\tau\bm{K}_v-\bm{p}_v}^\dagger
+
\hat{V}_{IL}
,
\end{equation}
where $E_{c}(\tau\bm{K}_c+\bm{p}_c)$ and $E_{v}(\tau\bm{K}_v+\bm{p}_v)$ are the conduction and valence electron dispersion, respectively.
At low energy, we can expand these bands near the valleys $\tau\bm{K}_n$ to quadratic order in $\bm{p}_n$, indicating:
\begin{equation}
\hat{\mathcal{H}}_0
\simeq
\sum_{\tau}
\int d^2\bm{s}
\hat{c}_{\tau}^\dagger(\bm{s})
\left[
\frac{\hat{\bm{p}}_{c,\tau}^2}{2m_{c}}
+
E_{c}^0
+
\mu
\right]
\hat{c}_{\tau}(\bm{s})
+
\hat{v}_{\tau}^\dagger(\bm{s})
\left[
\frac{\hat{\bm{p}}_{v,\tau}^2}{2m_{v}}
-
E_{v}^{0}
-
\mu
\right]
\hat{v}_{\tau}(\bm{s})
+
\hat{V}_{IL}
,\quad
\hat{\bm{p}}_{n,\tau} = -i\nabla_{\bm{s}} - \tau\bm{K}_{n}
,
\end{equation}
where $\hat{c}_{\tau}(\bm{s})$ and $\hat{v}_{\tau}(\bm{s})$ are Fourier transform of $\hat{c}_{\tau\bm{K}_c+\bm{k}_c}$ and $\hat{v}_{-\tau\bm{K}_v-\bm{k}_v}$ to the (monolayer) Wannier basis, with the in-plane site positions treated as a continuous two-dimensional variable $\bm{s}$ because we only care about length scales much larger than $|\bar{\bm{a}}|$.
Here $m_{c}$ and $m_{v}$ are the effective masses of conduction electron and valence \textit{hole}, respectively, and $E_{n}^0\equiv E_{n}(\tau\bm{K}_n)$ is the energy offset of each band and layer.

Further progress requires details of the interlayer coupling term.
For simplicity, we will consider heterobilayers such that an offset exists between bands of the two layers, which renders the interlayer coupling perturbative.
In other words, in these situations such coupling simply introduces moir\'e potential $\Delta_{c}(\bm{s})$ and $\Delta_{v}(\bm{s})$ to the charges, giving:
\begin{equation}
\label{eq:H_0}
\hat{\mathcal{H}}_0
=
\sum_{n,\tau}
\int d^2\bm{s}
\hat{\psi}_{\tau}^{(n)\dagger}(\bm{s})
h_n(\bm{s})
\hat{\psi}_{\tau}^{(n)}(\bm{s})
,\quad
h_n(\bm{s})
=
\frac{\hat{\bm{p}}_{n,\tau}^2}{2m_{n}}
+
\Delta_{n}(\bm{s})+(\delta_{n,c}-\delta_{n,v})(E_{n}^{0}+\mu)
,
\end{equation}
where we use the short-hand notation $\hat{\psi}_{\tau}^{(c)}(\bm{s})=\hat{c}_{\tau}(\bm{s})$ and $\hat{\psi}_{\tau}^{(v)}(\bm{s})=\hat{v}_{\tau}(\bm{s})$.
This expression reproduces the non-interacting term in Eq.~(1) of the main text.

\subsection{Interactions}
\label{sec:interaction}

We proceed to express the interactions $\hat{\mathcal{H}}_L$ and $\hat{\mathcal{H}}_T$ with the degrees of freedom at each band.
Before implementing Eq.~\eqref{eq:electron_field} on these terms, we note that the low energy physics of TMDs lies at momenta near the valleys, implying that we can implement approximations based on small $p_n$ and small $\delta\bm{K}_n\equiv\bm{K}_n- \bar{\bm{K}}$.
More specifically, this consideration allows for the $\bm{k}\cdot\bm{p}$ approximation~\cite{Haug2004}, which brings the Bloch function $u_{\tau\bm{K}_n+\bm{p}_n}^{n}(\bm{r})=e^{-i(\tau\bm{K}_n+\bm{p}_n)\cdot\bm{r}}\psi_{\tau\bm{K}_n+\bm{p}_n}^{n}(\bm{r})$ to the following expression:
\begin{equation}
\label{eq:k_dot_p_Bloch_func}
u_{\tau\bm{K}_n+\bm{p}_n}^{n}(\bm{r})
\simeq
u_{\tau\bar{\bm{K}}}^{n}(\bm{r})
+
\frac{i}{e}
(\tau\delta\bm{K}_n+\bm{p}_n)
\cdot
\sum_{n'}
\bm{d}_{\tau}^{n',n}
u_{\tau\bar{\bm{K}}}^{n'}(\bm{r})
,\quad
\bm{d}_{\tau}^{n,n'}
\equiv
-
e
(1-\delta_{n,n'})
\int
d^3\bm{r}
\psi_{\tau\bar{\bm{K}}}^{n\ast}(\bm{r})
\bm{r}
\psi_{\tau\bar{\bm{K}}}^{n'}(\bm{r})
,
\end{equation}
where the $(1-\delta_{n,n'})$ factor in the transition dipole matrix elements $\bm{d}_{\tau}^{n,n'}$ results from the parities of conduction and valence electronic wavefunctions at valley $\tau$, which corresponds to $|d_{z^2}\rangle$ and $|d_{x^2-y^2}\rangle+i\tau|d_{xy}\rangle$ orbitals~\cite{xiao2012coupled}.
In the following, we utilize Eq.~\eqref{eq:electron_field} and Eq.~\eqref{eq:k_dot_p_Bloch_func} to simplify the interactions.

We start by analyzing $\hat{\mathcal{H}}_L$, which are weighted integral of (normal-ordering of) $\hat{\psi}^\dagger(\bm{r}) \hat{\psi}(\bm{r}) \hat{\psi}^\dagger(\bm{r}') \hat{\psi}(\bm{r}')$. 
For simplicity, we focus on only two contributions from this interaction, depending on whether the two fermions in the local electron density operator $\hat{\psi}^\dagger(\bm{r}) \hat{\psi}(\bm{r})$ belong to the same or different bands after substituting Eq.~\eqref{eq:electron_field}.

We begin with the situation where the two fermions in $\hat{\psi}^\dagger(\bm{r}) \hat{\psi}(\bm{r})$ are intraband.
Such term would contain only the zeroth order term in Eq.~\eqref{eq:k_dot_p_Bloch_func} because the intraband transition dipole matrix elements are vanishing.
The remaining terms can be straightforwardly simplified by utilizing the orthogonality relation and (approximate) discrete translational invariance (at small twist and lattice mismatch) of Bloch wavefunctions.
Together with a transformation from Bloch to the Wannier states and treating the monolayer sites as a continuous variable, we find that the target contributions in $\hat{\psi}^\dagger(\bm{r}) \hat{\psi}(\bm{r})$ eventually become $\sim \hat{\rho}_{n}(\bm{s}) = \sum_\tau \hat{\psi}_{\tau}^{(n)\dagger}(\bm{s}) \hat{\psi}_{\tau}^{(n)}(\bm{s})$.
We therefore find the following expression for the corresponding sector of $\hat{\mathcal{H}}_L$:
\begin{equation}
\hat{\mathcal{V}}_e
=
\frac{e^2}{8\pi\epsilon}
\int
d^2\bm{s}d^2\bm{s}'
\frac{:\hat{\rho}_{c}(\bm{s})\hat{\rho}_{c}(\bm{s}'):}{|\bm{s}-\bm{s}'|}
-
\frac{2\hat{\rho}_{c}(\bm{s})\hat{\rho}_{v}(\bm{s}')}{\sqrt{(\bm{s}-\bm{s}')^2+z_{cv}^2}}
,
\end{equation}
where $z_{cv}$ is the distance between the two layers and the colons indicating normal-ordering of the operators in between.
Note that here we replace $\epsilon_0$ with the static permittivity of the material $\epsilon$ to account for dielectric screening.

In constrast, for the situation where the two fermions in $\hat{\psi}^\dagger(\bm{r}) \hat{\psi}(\bm{r})$ are interband, the contribution from zeroth order term in Eq.~\eqref{eq:k_dot_p_Bloch_func} vanishes because $\int
d^3\bm{r}\psi_{\tau\bar{\bm{K}}}^{c\ast}(\bm{r})\psi_{\tau'\bar{\bm{K}}}^{v}(\bm{r})=0$, whereas the first order corrections (which involves transition dipoles) are generally non-vanishing.
Dropping only higher order corrections and repeating the procedures in calculating intraband contributions, we find the target terms in $\hat{\psi}^\dagger(\bm{r}) \hat{\psi}(\bm{r})$ eventually becomes $\sim \frac{1}{e} \nabla_{\bm{s}} \cdot \hat{\bm{P}}^\dagger(\bm{s})$ and its Hermitian conjugate, where $\hat{\bm{P}}^\dagger(\bm{s}) = \sum_{\tau} \bm{d}_{\tau}^{cv} \hat{c}_{\tau}^{\dagger}(\bm{s}) \hat{v}_{\tau}^{\dagger}(\bm{s})$ is the electric polarization operator.
The resulting operator from $\hat{\mathcal{H}}_L$ is the longitudinal dipole-dipole interaction, which has the following expression:
\begin{equation}
\hat{\mathcal{V}}_{d,L}
=
\frac{1}{\epsilon_0}
\int
d^2\bm{s}d^2\bm{s}'
\hat{\bm{P}}^\dagger(\bm{s})
\cdot
\left[
\nabla_{\bm{s}}
\otimes
\nabla_{\bm{s}'}
\frac{1}{|\bm{s}-\bm{s}'|}
\right]
\cdot
\hat{\bm{P}}(\bm{s}')
.
\end{equation}

We proceed with analyzing $\hat{\mathcal{H}}_T$, which could be turned into the following form after a canonical transformation~\footnote{Here we drop a term that is quartic in the tranverse electric polarization, which is expected to only lead to an overall energy shift to the emitters. See Ref.~\cite{steck2007quantum}.}:
\begin{equation}
\label{eq:H_T}
\hat{\mathcal{H}}_T
\simeq
-
\int d^2\bm{s}
\left[
\hat{\bm{P}}^\dagger(\bm{s})
\cdot
\hat{\bm{E}}_T\left(\bm{s}\right)
+
\mathrm{H.c.}
\right]
+
\hat{\mathcal{H}}_T^R
,\quad
\hat{\mathcal{H}}_T^R
=
\frac{\epsilon_0}{2}
\int d^3\bm{r}
\hat{\bm{E}}_T^\dagger(\bm{r})
\hat{\bm{E}}_T(\bm{r})
+
c^2
\hat{\bm{B}}^\dagger(\bm{r})
\hat{\bm{B}}(\bm{r})
,
\end{equation}
where $c$ is the speed of light, and $\hat{\bm{E}}_T(\bm{r})$ and $\hat{\bm{B}}(\bm{r})$ are the transverse electric and magnetic field operators, respectively, evaluated at position $\bm{r}$.
To obtain a quartic fermion interaction from this coupling, we integrate out the gauge fields following standard diagrammatic technique~\cite{abrikosov2012methods}.
This requires the propagator of transverse electric field such that the target vertex involves retardation effect.
We address this by assuming that the relevant photon fluctuations are near-resonant to the target pair state, whose energy is denoted as $\omega_p$.
The resulting interaction has the following transverse dipole-dipole form:
\begin{equation}
\hat{\mathcal{V}}_{d,T}
=
(-i)
\int
d^2\bm{s} d^2\bm{s}'
\hat{\bm{P}}^\dagger(\bm{s})
\cdot
\int_{-\infty}^\infty dt
\langle 
\hat{\bm{E}}_T(\bm{s})
e^{i(\omega_p-\hat{\mathcal{H}}_T^R)t}
\hat{\bm{E}}_T^\dagger(\bm{s}')
\rangle_{\mathrm{vac}}
\cdot
\hat{\bm{P}}(\bm{s}')
,
\end{equation}
where $\langle ... \rangle_{\mathrm{vac}}$ denotes the vacuum expectation value of the operator in between.

The longitudinal and transverse dipole-dipole interactions can be combined into~\cite{steck2007quantum,perczel2017photonic}:
\begin{equation}
\hat{\mathcal{V}}_d
=
\hat{\mathcal{V}}_{d,L}
+
\hat{\mathcal{V}}_{d,T}
=
\frac{\omega_p^2}{c^2\epsilon_0}
\int
d^2\bm{s} d^2\bm{s}'
\hat{\bm{P}}^\dagger(\bm{s})
\cdot
\bm{\mathcal{G}}\left(\omega_p,\bm{s}-\bm{s}'\right)
\cdot
\hat{\bm{P}}(\bm{s}')
,
\end{equation}
where $\mathcal{G}_{\alpha,\beta}(ck,\bm{r})$ denotes the ($\alpha,\beta\in x,y,z$ components of) dyadic Green's tensor:
\begin{equation}
\label{eq:dyadic_Green_tensor}
\mathcal{G}_{\alpha,\beta}(ck,\bm{r})
=
-\frac{e^{ikr}}{4\pi r}
\left[
\left(
1+\frac{i}{kr}-\frac{1}{(kr)^2}
\right)
\delta_{\alpha,\beta}
+
\left(
-1-\frac{3i}{kr}+\frac{3}{(kr)^2}
\right)
\frac{r_\alpha r_\beta}{r^2}
\right]
+
\frac{\delta_{\alpha,\beta}\delta^{(3)}(\bm{r})}{3k^2}
.
\end{equation}
Note that unlike the static interaction, here we neglect the dielectric screening at frequency $\omega_p$ such that the permittivity here is still $\epsilon_0$.
This is valid if $\omega_p$ is off-resonant with other excitations (except the target pair).
Another important point is that the interaction $\hat{\mathcal{V}}_d$ influences the dynamics of an electron-hole pair even in the dilute regime and is physically distinct from the Hubbard-type electrostatic repulsion between pairs~\cite{park2023dipole}, which becomes significant only at high densities of these excitations, a regime that is not the focus of this work.

Before moving on, we comment on a few subtleties regarding the \textit{interlayer} terms in the dipole-dipole interaction.
First, renormalization from the interlayer coupling should also be considered in the transition dipole matrix elements on top of its expression in Eq.~\eqref{eq:k_dot_p_Bloch_func}.
More specifically, contributions from a two-step process in which interlayer coupling occurs after a (virtual) creation of intralayer electron-hole pair is non-negligible~\cite{yu2015anomalous}.
Therefore, for interlayer transitions, we will not compute $\bm{d}_{\tau}^{cv}$ using Eq.~\eqref{eq:k_dot_p_Bloch_func} but instead quote the results in literature~\cite{yu2015anomalous, barre2022optical}.
Second, taking the continuum limit of the monolayer lattice positions in the dipole-dipole interaction is nontrivial because, strictly speaking, the displacement vector between the sites at different layer is position dependent due to twisting or lattice mismatch~\cite{wu2018theory}, which could lead to a position-dependent dipole matrix element~\cite{tran2019evidence}.
Nevertheless, here we assume that the spatial fluctuations of electron-hole states of interest is small compared to the scale at which such displacement becomes non-negligible (which is characterized by the moir\'e period $a_M$)~\cite{tran2019evidence,karni2022structure,park2023dipole}, validating our continuum approximation.

\section{The effective moir\'e superlattice model}

In this section, we derive the superlattice model for optical excitations in the lowest energy manifold and for doped electrons from the two band Hamiltonian described in the main text, starting from the simple case without electron doping.

\subsection{Undoped bilayers}

In the absence of doping, the optical excitations are given by electron-hole pairs.
We specifically focus on the lowest energy pairs in the dilute regime, which allows us to project the two band model $\hat{\mathcal{H}}$ into the single excitation basis $\hat{x}_{\bm{R},\tau}^\dagger|\mathrm{vac}\rangle$, where $|\mathrm{vac}\rangle$ denotes the vacuum state and the creation operator of a pair in the lowest energy moir\'e-Wannier basis is:
\begin{equation}
\hat{x}_{\bm{R},\tau}^\dagger
=
\int
d^2\bm{s}_c
d^2\bm{s}_v
w_{\bm{R}}(\bm{s}_c,\bm{s}_v)
\hat{c}_\tau^{\dagger}(\bm{s}_c)
\hat{v}_\tau^{\dagger}(\bm{s}_v)
.
\end{equation}
Here $w_{\bm{R}}(\bm{s}_c,\bm{s}_v)$ is the Wannier orbital centered at supersite $\bm{R}$, with $\bm{s}_c$ and $\bm{s}_v$ being the electron and hole in-plane coordinates, respectively.
Note that $w_{\bm{R}}(\bm{s}_c,\bm{s}_v)$ takes the functional form $f(\bm{s}_c-\bm{R},\bm{s}_v-\bm{R})$ due to superlattice periodicity, and at this point we assume it is valley independent.
This projection yields:
\begin{equation}
\hat{\mathcal{H}}
\to
\hat{H}
=
\left(
\omega_{\mathrm{ex}}-\frac{i\gamma}{2}
\right)
\sum_{\bm{R},\tau}
\hat{x}_{\bm{R},\tau}^\dagger
\hat{x}_{\bm{R},\tau}
+
\xi
\sum_{\bm{R},\tau}
\hat{x}_{\bm{R},\tau}^\dagger
\hat{x}_{\bm{R},-\tau}
-
\sum_{\bm{R}\neq\bm{R}'}
\sum_{\tau,\tau'}
t_{\bm{R},\bm{R}'}^{\tau,\tau'}
\hat{x}_{\bm{R},\tau}^\dagger
\hat{x}_{\bm{R}',\tau'}
,
\end{equation}
where the on-site energy integrals are:
\begin{equation}
\omega_{\mathrm{ex}}
=
\int
d^2\bm{s}_c
d^2\bm{s}_v
w_{\bm{R}}^{\ast}(\bm{s}_c,\bm{s}_v)
h_{\mathrm{ex}}(\bm{s}_c,\bm{s}_v)
w_{\bm{R}}(\bm{s}_c,\bm{s}_v)
+
\frac{\omega_{\mathrm{ex}}^2|d|^2}{c^2\epsilon_0}
\bm{e}_{\tau}^\ast
\cdot
\mathrm{Re}
[
\bm{\mathcal{G}}^{\mathrm{ex}}\left(0\right)
]
\cdot
\bm{e}_{\tau}
,
\end{equation}
\begin{equation}
\gamma
=
-
\frac{\omega_{\mathrm{ex}}^2|d|^2}{c^2\epsilon_0}
\bm{e}_{\tau}^\ast
\cdot
2
\mathrm{Im}
[
\bm{\mathcal{G}}^{\mathrm{ex}}\left(0\right)
]
\cdot
\bm{e}_{\tau}
,\quad
\xi
=
\frac{\omega_{\mathrm{ex}}^2|d|^2}{c^2\epsilon_0}
\bm{e}_{\tau}^\ast
\cdot
\mathrm{Re}
[
\bm{\mathcal{G}}^{\mathrm{ex}}\left(0\right)
]
\cdot
\bm{e}_{-\tau}
,
\end{equation}
where we use $\bm{d}_{\tau}^{cv} = d\bm{e}_\tau^\ast$ as well as the shorthand notations $h_{\mathrm{ex}}(\bm{s}_c,\bm{s}_v)=\sum_nh_n(\bm{s}_n)-\frac{e^2}{4\pi\epsilon\sqrt{(\bm{s}_c-\bm{s}_v)^2+z_{cv}^2}}$ and:
\begin{equation}
\bm{\mathcal{G}}^{\mathrm{ex}}\left(\bm{R}-\bm{R}'\right)
=
\int
d^2\bm{s}
d^2\bm{s}'
w_{\bm{R}}^{\ast}(\bm{s},\bm{s})
\bm{\mathcal{G}}\left(\omega_{\mathrm{ex}};\bm{s}-\bm{s}'\right)
w_{\bm{R}'}^{\mathrm{ex}}(\bm{s}',\bm{s}')
.
\end{equation}
In addition, there are tunneling terms from the intrinsic charge dynanics $t_{\bm{R},\bm{R}'}^{\mathrm{int}}$ and from the dipole-dipole interaction.
Combining the two terms gives:
\begin{equation}
t_{\bm{R},\bm{R}'}^{\tau,\tau'}
=
t_{\bm{R},\bm{R}'}^{\mathrm{int}}
\delta_{\tau,\tau'}
-
\frac{\omega_{\mathrm{ex}}^2|d|^2}{c^2\epsilon_0}
\bm{e}_{\tau}^\ast
\cdot
\bm{\mathcal{G}}^{\mathrm{ex}}\left(\bm{R}-\bm{R}'\right)
\cdot
\bm{e}_{\tau'}
,\quad
t_{\bm{R},\bm{R}'}^{\mathrm{int}}
=
-
\int
d^2\bm{s}_c
d^2\bm{s}_v
w_{\bm{R}}^{\ast}(\bm{s}_c,\bm{s}_v)
h_{\mathrm{ex}}(\bm{s}_c,\bm{s}_v)
w_{\bm{R}'}^{\mathrm{ex}}(\bm{s}_c,\bm{s}_v)
.
\end{equation}

To further simplify the model, we consider the regime where $a_M$ is controlled (by twisting angle) to be sufficiently large, allowing for the approximations below.
First, this consideration implies the width of $w_{\bm{R}}(\bm{s},\bm{s})$ in $\bm{s}$, denoted as $a_W$, is also large.
Under this circumstance, the on-site term given by the Green's tensor, which roughly scale with $\sim\frac{d^2}{\epsilon a_W^3}$, is suppressed compared to the center-of-mass sector of $\int d^2\bm{s}_cd^2\bm{s}_vw_{\bm{R}}^{\ast}(\bm{s}_c,\bm{s}_v)h_{\mathrm{ex}}(\bm{s}_c,\bm{s}_v)w_{\bm{R}}(\bm{s}_c,\bm{s}_v)$, which scales with $\sim\frac{1}{2(m_c+m_v)a_W^2}$~\footnote{Typically $a_W\ll a_M$ such that one can expand the moir\'e potential to second order in space near $\bm{R}$.}.
This comparison indicates that $\xi$ and $\gamma$ only appear as high order corrections to the eigenvalues of the model.
Second, a large $a_M$ indicates the validity of tight-binding approximation, suggesting that off-site terms are perturbative compared to on-site ones.
Combining these considerations, the zeroth order energy from $\hat{H}$ is simply $\omega_{\mathrm{ex}}$, and the functional form of $w_{\bm{R}}(\bm{s}_c,\bm{s}_v)$ can therefore be determined by this quantity (as by definition it is the wavefunction of the lowest energy manifold), subject to the constraint that it only depends on $\bm{s}_c-\bm{R}$ and $\bm{s}_v-\bm{R}$.
The resulting wavefunction is independent of $\tau$, consistent with the assumption made at the very beginning of this section.
In addition, it is a representation of C3 rotation, suggesting that the first order eigenvalue correction from $\xi$ vanishes, and therefore, one can neglect its contribution in $\hat{H}$ and recover the superlattice model presented in the main text.

\subsubsection{Details of the dyatic Green's tensor}
\label{eq:exciton_Greens_tensor}

Before moving on to doped bilayers, we elaborate on details of $\bm{\mathcal{G}}^{\mathrm{ex}}\left(\bm{R}-\bm{R}'\right)$, which depends on the moir\'e-Wannier function of exciton $w_{\bm{R}}(\bm{s},\bm{s})$.
Simplification of this Green's tensor therefore requires the analytical form of this wavefunction, which we assume to be the following form for simplicity~\cite{huang2023non}:
\begin{equation}
\label{eq:w_ansatz}
w_{\bm{R}}(\bm{s}_c,\bm{s}_v)
=
\frac{2\sqrt{2}}{\pi a_Ba_W}
\exp\left[-\frac{(\bm{s}_{\mathrm{ex}}-\bm{R})^2}{2a_W^2}-\frac{2|\bm{s}_c-\bm{s}_v|}{a_B}\right]
,\quad
\bm{s}_{\mathrm{ex}}
=
\frac{m_c\bm{s}_c+m_v\bm{s}_v}{m_c+m_v}
,
\end{equation}
where $a_B$ characterizes the electron-hole relative distance (and recall that $a_W$ is the center-of-mass spatial fluctuation).
This expression is strictly valid if (a) $a_W\ll a_M$ such that quadratic expansions to the center-of-mass potential terms are applicable, and (b) the coupling between center-of-mass and relative coordinates provided by the moir\'e potentials are suppressed.
Utilizing this expression, the integrated Green's tensor becomes:
\begin{equation}
\mathcal{G}_{\alpha,\beta}^{\mathrm{ex}}\left(\bm{R}-\bm{R}'\right)
=
\frac{32a_W^2}{a_B^2}
\int\frac{d^3\bm{p}}{(2\pi)^3}
e^{i\bm{p}\cdot(\bm{R}-\bm{R}')}
e^{-p^2a_W^2}
\frac{k^2\delta_{\alpha,\beta}-p_\alpha p_\beta}{k^2(k^2-p^2)}
\bigg|_{k = \frac{\omega_{\mathrm{ex}}}{c}}
,
\end{equation}
where an ultraviolet regulator to $\bm{p}$ is introduced naturally by $a_W$.
This yields the following on-site contribution:
\begin{equation}
\mathcal{G}_{\alpha,\beta}^{\mathrm{ex}}\left(0\right)
=
\delta_{\alpha,\beta}
\frac{16ka_W^2}{3\pi a_B^2}
\left[
\frac{\mathrm{erfi}(ka_W)-i}{e^{k^2a_W^2}}
-
\frac{(ka_W)^2-\frac{1}{4}}{\sqrt{\pi}(ka_W)^3}
\right]
\bigg|_{k = \frac{\omega_{\mathrm{ex}}}{c}}
,\quad
\mathrm{erfi}(ka_W)=\frac{2}{\sqrt{\pi}}\int_0^{ka_W}dy\exp(y^2)
,
\end{equation}
which directly indicates:
\begin{equation}
\gamma
\simeq
\frac{32a_W^2}{a_B^2}
\gamma_{cv}
,\quad
\gamma_{cv}
=
\frac{|d|^2\omega_{\mathrm{ex}}^3}{3\pi\epsilon_0 c^3}
,
\end{equation}
where we take the approximation $e^{k^2a_W^2}\simeq1$ as the array is subwavelength.
Note that here $\gamma_{cv}$ is the Wigner-Weisskopf spontaneous emission rate for conduction-valence band (of monolayers) transition that is independent of $a_W$ and $a_B$.
In contrast, $\gamma$ is the bare radiative decay rate of a moir\'e exciton, which is corrected by a prefactor $\frac{32a_W^2}{a_B^2}$ set by length scales of the pair wavefunction. 
Notably, $a_W$ generally depends on $a_M$, indicating that $\gamma$ varies with twisting angle.

Finally, for later convenience, we discuss two different Fourier transforms of the Green's tensor, which brings $\bm{\mathcal{G}}^{\mathrm{ex}}\left(\bm{R}\right)$ and $\bm{\mathcal{G}}\left(ck,\bm{r}-\bm{R}\right)$ into the center-of-mass Bloch momentum space labeled by $\bm{Q}$.
Note that $\bm{Q}$ is Fourier conjugate of the emergent lattice vectors, denoted as $\bm{L}$, which is different from $\bm{R}$ if $\{\bm{R}\}$ defines a non-Bravais lattices.
The first transformation reads:
\begin{equation}
\label{eq:g_sum_G}
\sum_{\bm{L}\neq\bm{b}'-\bm{b}}
\bm{\mathcal{G}}^{\mathrm{ex}}(\bm{L}+\bm{b}-\bm{b}')
e^{-i\bm{Q}\cdot(\bm{L}+\bm{b}-\bm{b}')}
=
\frac{1}{\mathcal{A}}
\sum_{\bm{G}}
e^{i\bm{G}\cdot(\bm{b}-\bm{b}')}
\tilde{\bm{\mathcal{G}}}^{\mathrm{ex}}(\bm{Q}+\bm{G})
-
\delta_{\bm{b}',\bm{b}}
\bm{\mathcal{G}}^{\mathrm{ex}}(0)
,
\end{equation}
where we replace $\bm{R}=\bm{L}+\bm{b}$ with $\bm{b}$ being sublattice vectors for non-Bravais lattices.
Here $\mathcal{A}$ is the unit emergent cell area, $\bm{G}$ denotes reciprocal lattice vectors, and:
\begin{equation}
\label{eq:tilde_mathcal_G}
\tilde{\mathcal{G}}_{\alpha,\beta}^{\mathrm{ex}}(\bm{q})
=
\frac{32a_W^2}{a_B^2}
\left(
\delta_{\alpha,\beta}
-
\frac{c^2q_\alpha q_\beta}{\omega_{\mathrm{ex}}^2}
\right)
I(\omega_{\mathrm{ex}},q)
,\quad
\forall
\alpha,\beta\neq z
,\;
\bm{q}\cdot\bm{e}_z=0
,
\end{equation}
with: 
\begin{equation}
\label{eq:I_ck_q}
I(ck,q)
=
\frac{
e^{-k^2a_W^2}
\left[
\mathrm{erfi}\left(\sqrt{k^2-q^2}a_W\right)-i
\right]
}{2\sqrt{k^2-q^2}}
,
\end{equation}
where $\sqrt{k^2-q^2}$ has non-negative real and imaginary parts.
This type of Fourier transformation is useful upon evaluation of the low-$Q$ Hamiltonians in Section~\ref{sec:low_kB_H}.
In contrast, the second transformation reads:
\begin{equation}
\label{eq:g_r_ck_kB_b}
\sum_{\bm{L}}
\bm{\mathcal{G}}(ck,\bm{s}+z\bm{e}_z-\bm{L}-\bm{b})
e^{i\bm{Q}\cdot(\bm{L}+\bm{b})}
\simeq
\bm{g}(k,\bm{Q})
e^{i\bm{Q}\cdot\bm{s}-i\sqrt{k^2-Q^2}z}
,\quad
\forall z<0,\;|z|\gg a_M
,
\end{equation}
\begin{equation}
g_{\alpha,\beta}(k,\bm{Q})
=
-
\frac{i}{2\mathcal{A}}
\frac{
1
}{\sqrt{k^2-Q^2}}
\left[
\delta_{\alpha,\beta}
-
\frac{Q_{\alpha}Q_{\beta}}{k^2}
\right]
.
\end{equation}
where only zeroth order diffraction is kept as others are exponentially suppressed by a factor $\sim \exp(-|z|/a_M)$, which are negligible in the far-field limit.
This type of transformation is utilized to evaluate the scattering matrix of light in Section~\ref{sec:light_scatter}.

\subsection{Doped bilayers}
\label{sec:doped}

We proceed by generalizing the superlattice model to bilayers with Wigner crystals consisting of doped electrons.
Before doing this, we briefly review how those charge orders emerge from the two band model.
To begin with, the doped electrons fall into the first conduction moir\'e-Wannier state (which is split from the conduction band by moir\'e potential) given by the following creation operator:
\begin{equation}
\label{eq:f_R_tau}
\hat{f}_{\bm{R},\tau}^\dagger
=
\int
d^2\bm{s}
w_{\bm{R}}^c(\bm{s})
\hat{c}_\tau^\dagger(\bm{s})
,\quad
w_{\bm{R}}^c(\bm{s})
=
\frac{1}{\sqrt{N}}
\sum_{\bm{Q}}
e^{-i\bm{Q}\cdot\bm{R}}
\phi_{\bm{Q}}^c(\bm{s})
,\quad
h_c(\bm{s})
\phi_{\bm{Q}}^c(\bm{s})
=
E_{c,\bm{Q}}
\phi_{\bm{Q}}^c(\bm{s})
,
\end{equation}
where $\bm{R}$ labels the supersites and $N=\sum_{\bm{R}}1$.
Here $w_{\bm{R}}^c(\bm{s})$ and $\phi_{\bm{Q}}^c(\bm{s})$ are the lowest moir\'e-Wannier and moir\'e-Bloch wavefunctions, respectively. 
The latter is labeled by superlattice momentum $\bm{Q}$ because the corresponding eigenvalue problem (with eigenvalue $E_{c,\bm{Q}}$) has moir\'e periodicity. 
Projecting the two band model into this fermionic degrees of freedom and include only terms depending on $\hat{f}_{\bm{R},\tau}^\dagger\hat{f}_{\bm{R},\tau}$ in the interaction, we find the following superlattice model for doped electrons:
\begin{equation}
\hat{H}_f
=
E_f
\sum_{\bm{R},\tau}
\hat{f}_{\bm{R},\tau}^\dagger
\hat{f}_{\bm{R},\tau}
-
\sum_\tau
\sum_{\bm{R},\bm{R}'}
t_{\bm{R}',\bm{R}}^f
\hat{f}_{\bm{R}',\tau}^\dagger
\hat{f}_{\bm{R},\tau}
+
\frac{1}{2}
\sum_{\tau,\tau'}
\sum_{\bm{R},\bm{R}'}
U_f(\bm{R}-\bm{R}')
\hat{f}_{\bm{R},\tau}^\dagger
\hat{f}_{\bm{R}',\tau'}^\dagger
\hat{f}_{\bm{R}',\tau'}
\hat{f}_{\bm{R},\tau}
,
\end{equation}
where $E_f=\int d^2\bm{s}w_{\bm{R}}^{c\ast}(\bm{s}) h_c(\bm{s}) w_{\bm{R}}^c(\bm{s})$, $t_{\bm{R}',\bm{R}}^f=-\int d^2\bm{s}w_{\bm{R}'}^{c\ast}(\bm{s}) h_c(\bm{s})w_{\bm{R}}^c(\bm{s})$, and $U_f(\bm{R}-\bm{R}')=\int d^2\bm{s} d^2\bm{s}'\frac{e^2|w_{\bm{R}}^c(\bm{s})w_{\bm{R}'}^c(\bm{s}')|^2}{4\pi\epsilon|\bm{s}-\bm{s}'|}$.
It has been shown that this superlattice model with parameters from TMD bilayers could realize Wigner crystal states at certain fractional fillings of doped electrons~\cite{pan2020quantum}.

Next, we consider adding electron-hole pairs into these charge orders.
We again stick to the large $a_M$ assumption such that the zeroth order eigenvalues of the full model can be determined by an unit supercell problem.
In this situation, the coexistence of one pair with localized fermions provides two possibilities, depending on whether the pair lies in a cell with a doped electron, giving a three-particle and two-particle intra-cell state, respectively.
Their energies ($\omega_t$ and $\omega_{\mathrm{ex}}$) possess a difference of which sign depends on whether the generated electron-hole pair is intralayer or interlayer~\cite{ciorciaro2023kinetic,gao2024excitonic}.
We specifically focus on the interlayer scenario, giving $\omega_t-\omega_{\mathrm{ex}}\simeq30$meV for zero twist WSe\textsubscript{2}/WS\textsubscript{2}~\cite{gao2024excitonic}.
This offset is large enough to suppress the coupling between these sectors~\footnote{Such coupling results from hopping of the pair state. The nearest-neighbor intra-valley tunneling from the Green's tensor is estimated to be $\sim\frac{|d|^2}{\epsilon a_M^2}$, which is roughly $\simeq 1.6$meV utilizing standard values of $d=0.5$e$\cdot$nm and dielectric constant $5$~\cite{barre2022optical,park2023dipole}, which is much smaller than $\omega_t-\omega_{\mathrm{ex}}$}, suggesting that the lowest energy pairs are simply the ones living in supercells without doped electrons.
The superlattice Hamiltonian discussed in the previous section therefore applies even in the presence of Wigner crystal state of doped electrons, except that $\{\bm{R}\}$ preoccupied by these charges have to be projected out.

\section{Low momentum Hamiltonians near the moir\'e Brillouin zone center}
\label{sec:low_kB_H}

In this section, we discuss the low-$Q$ Hamiltonians for different lattice structures.
We specifically focus on Bravais lattices (triangular and rectangular arrays) where sublattice indices are trivial such that the low-$Q$ Hamiltonian matrix elements can be denoted as $h_{\tau,\tau'}(\bm{Q})$.

We begin by evaluating the Bloch Hamiltonian exactly at $\bm{Q}=0$, which reads:
\begin{equation}
h_{\tau,\tau'}(0)
=
\left(
\omega_{\mathrm{ex}}
+
\tau \mu_BB
-i
\frac{\gamma}{2}
\right)
\delta_{\tau,\tau'}
+
\frac{|d|^2\omega_{\mathrm{ex}}^2}{2c^2\epsilon_0}
\left[
\delta_{\tau,\tau'}
\left(
h_{xx}
+
h_{yy}
\right)
+
(1-\delta_{\tau,\tau'})
\left(
h_{xx}
-
h_{yy}
-2i\tau
h_{xy}
\right)
\right]
,
\end{equation}
where $\mu_BB$ denotes the Zeeman splitting from out-of-plane magnetic field and $h_{\alpha,\beta}\equiv\sum_{\bm{L}\neq0}
\mathcal{G}_{\alpha,\beta}^{\mathrm{ex}}(\bm{L})$.
Notably, if $\{\bm{L}\}=\{\check{O}\bm{L}\}$ and $\mathcal{G}_{\alpha,\beta}^{\mathrm{ex}}(\bm{L})=\mathcal{G}_{\alpha,\beta}^{\mathrm{ex}}(\check{O}\bm{L})$, where $\check{O}$ denotes the operations of C$_3$ point group, we find $h_{xx}=h_{yy}$ and $h_{xy}=0$, indicating that $h_{\tau,\tau'}(0)\sim\delta_{\tau,\tau'}$.
This holds for triangular lattice but is not applicable for rectangular lattice, which is consistent with the fact that the former is doubly degenerate while the latter are split at $\mu_BB=0$ (see Main text and Section~\ref{sec:supplementary_data}).

With the zeroth order correction established, we proceed to the first order perturbation $\bm{Q}\cdot\nabla_{\bm{Q}}h_{\tau,\tau'}(\bm{Q})$.
Notably, all terms therein are associated with the summation $\sum_{\bm{L}\neq0}\bm{L}
\mathcal{G}_{\alpha,\beta}^{\mathrm{ex}}(\bm{L})$, which vanishes if $\{\bm{L}\}=\{\check{O}'\bm{L}\}$ and $\mathcal{G}_{\alpha,\beta}^{\mathrm{ex}}(\bm{L})=\mathcal{G}_{\alpha,\beta}^{\mathrm{ex}}(\check{O}'\bm{L})$, where $\check{O}'$ denotes the operations of C$_2$ point group.
Both triangular and rectangular lattices satisfy these conditions and therefore their first order corrections are zero.

Next, we discuss the second order corrections, which reads:
\begin{equation}
h_{\tau,\tau'}(\bm{Q})
-
h_{\tau,\tau'}(0)
\simeq
\frac{|d|^2\omega_{\mathrm{ex}}^2}{2c^2\epsilon_0}
\left[
\delta_{\tau,\tau'}
\left(
\delta^2h_{xx}
+
\delta^2h_{yy}
\right)
+
(1-\delta_{\tau,\tau'})
\left(
\delta^2h_{xx}
-
\delta^2h_{yy}
-2i\tau
\delta^2h_{xy}
\right)
\right]
,
\end{equation}
where:
\begin{equation}
\delta^2h_{\alpha,\beta}
=
\frac{1}{\mathcal{A}}
\sum_{\gamma,\delta}
Q_{\gamma}Q_{\delta}
\sum_{\bm{G}}
[
\partial_{q_{\gamma}}
\partial_{q_{\delta}}
\tilde{\mathcal{G}}_{\alpha,\beta}^{\mathrm{ex}}(\bm{q})
]_{\bm{q}\to\bm{G}}
.
\end{equation}
Rigourously speaking, one needs to evaluate the second derivative and perform the full $\bm{G}$ summation to get this correction.
Nevertheless, the $\bm{G}$ summand is exponentially suppressed with $Ga_W\sim \frac{a_W}{a_M}$ because it contains $I(\omega_{\mathrm{ex}},G)$, see Eq.~\eqref{eq:I_ck_q}, such that one only needs to include terms with $Ga_W\ll1$.
Here we proceed with simply the $G=0$ term, which yields:
\begin{equation}
h_{\tau,\tau'}(\bm{Q})
\simeq
h_{\tau,\tau'}(0)
+
iJ
(1-\delta_{\tau,\tau'})
(\bm{e}_\tau^\ast\cdot\bm{Q})^2
,\quad
J
=
\frac{3\pi c^4}{\mathcal{A}\omega_{\mathrm{ex}}^4}
\gamma
.
\end{equation}
Using the fact that $h_{\tau,-\tau}(0)=0$ for triangular lattice and going to polar coordinates for $\bm{Q}$, we recover the low-$Q$ model presented in the main text.

\section{Scattering theory of light}
\label{sec:light_scatter}

In this section, we review the scattering theory of light incident upon a two-dimensional lattice~\cite{shahmoon2017cooperative}, and apply it to obtain the reflection coefficient dressed by the collective properties of moir\'e excitons.
We start with the following integral solution to Maxwell equation for monochromatic electric field at frequency $ck$, $\bm{E}(\bm{r})e^{-ickt}$:
\begin{equation}
\label{eq:Maxwell_sol}
\nabla\times\nabla\times\bm{E}(\bm{r})
-
k^2\bm{E}(\bm{r})
=
\frac{k^2}{\epsilon_0}
\bm{P}(\bm{r})
\;
\to
\;
E_\tau(\bm{r})
=
E_{0,\tau}(\bm{r})
-
\frac{k^2}{\epsilon_0}
\sum_{\tau'}
\int d^3\bm{r}'
\mathcal{G}_{\tau,\tau'}(ck, \bm{r}-\bm{r}')
\cdot
P_{\tau'}(\bm{r}')
,
\end{equation}
where $E_\tau(\bm{r})$ is the $\bm{e}_\tau$ component of $\bm{E}(\bm{r})$, $\sum_\tau\bm{e}_\tau E_{0,\tau}(\bm{r})e^{-ickt}$ is the input field, $\mathcal{G}_{\tau,\tau'}(ck, \bm{r}-\bm{r}') = \bm{e}_\tau^\ast\cdot\bm{G}(ck, \bm{r}-\bm{r}')\cdot\bm{e}_{\tau'}$, and $\bm{P}(\bm{r})\equiv\langle\hat{\bm{P}}(\bm{r})\rangle = \sum_\tau\bm{e}_\tau P_{\tau}(\bm{r})$ is the (classical) electric polarization.
Upon suppression of the spatial extension of the pair wavefunction in the out-of-plane direction, projection to the lowest excitonic manifold, and dropping of the counter-rotating term $\sim \langle\hat{x}_{\bm{R},\tau}^\dagger\rangle$, the polarization vector becomes $\bm{P}(\bm{s}+z\bm{e}_z)\simeq\delta(z)d^\ast\sum_{\bm{R},\tau}\bm{e}_{\tau}w_{\bm{R}}(\bm{s},\bm{s})\langle\hat{x}_{\bm{R},\tau}\rangle$.

Further progress requires evaluation of $\langle\hat{x}_{\bm{R},\tau}\rangle$, whose dynamics are described by the following equation of motion in the weak drive limit (c.f.~Eq.~(4) in the main text):
\begin{equation}
ck
\langle
\hat{x}_{\bm{R},\tau}
\rangle
=
\sum_{\bm{R}',\tau'}
H_{\bm{R},\bm{R}'}^{\tau,\tau'}
\langle
\hat{x}_{\bm{R}',\tau'}
\rangle
-
\frac{4\sqrt{2}da_W}{a_B}
E_{0,\tau}(\bm{R})
,\quad
H_{\bm{R},\bm{R}'}^{\tau,\tau'}
=
\left(\omega_{\mathrm{ex}}+\tau\mu_BB-\frac{i\gamma}{2}\right)\delta_{\bm{R},\bm{R}'}\delta_{\tau,\tau'}
-
(1-\delta_{\bm{R},\bm{R}'})
t_{\bm{R},\bm{R}'}^{\tau,\tau'}
,
\end{equation}
where $H_{\bm{R},\bm{R}'}^{\tau,\tau'}$ is the matrix element of the superlattice Hamiltonian presented in the main text (with magnetic field). 
As we assume the drive does not vary significantly at the scale of $z_{cv}$ and $a_W$, we set its $z$ and $\bm{s}$ arguments as zero and $\bm{R}$, respectively.
We also utilize Eq.~\eqref{eq:w_ansatz} to set $\int d^2\bm{s}w_{\bm{R}}^{\ast}(\bm{s},\bm{s})=\frac{4\sqrt{2}a_W}{a_B}$.
The above equation can be simply solved by matrix inversion:
\begin{equation}
\langle
\hat{x}_{\bm{R},\tau}
\rangle
=
-
\frac{4\sqrt{2}da_W}{a_B}
\sum_{\bm{R}',\tau'}
D_{\bm{R},\bm{R}'}^{\tau,\tau'}(ck)
E_{0,\tau'}(\bm{R}')
,\quad
\sum_{\bm{R}_1,\tau_1}
\left[
ck
\delta_{\bm{R},\bm{R}_1}
\delta_{\tau,\tau_1}
-
H_{\bm{R},\bm{R}_1}^{\tau,\tau_1}
\right]
D_{\bm{R}_1,\bm{R}_2}^{\tau_1,\tau_2}(ck)
=
\delta_{\bm{R},\bm{R}_2}
\delta_{\tau,\tau_2}
,
\end{equation}
where $D_{\bm{R},\bm{R}'}^{\tau,\tau'}(ck)$ is the exciton propagator.
Plugging this expression back to Eq.~\eqref{eq:Maxwell_sol} yields an expression of $\bm{E}(\bm{r})$ involving the integral $\int d^2\bm{s}'\bm{\mathcal{G}}(ck, \bm{r}-\bm{s}')w_{\bm{R}}(\bm{s}',\bm{s}')$.
To simplify it, we assume that the field $\bm{E}(\bm{r})$ is eventually detected at far out-of-plane distance, indicating that $\bm{\mathcal{G}}(ck, \bm{r}-\bm{s}')\simeq \bm{\mathcal{G}}(ck, \bm{r}-\bm{s}'-\delta\bm{s}')$ for small $\delta\bm{s}'$, which validates the following approximation:
\begin{equation}
\int d^2\bm{s}'
\bm{\mathcal{G}}(ck, \bm{r}-\bm{s}')
w_{\bm{R}}(\bm{s}',\bm{s}')
\simeq
\frac{4\sqrt{2}a_W}{a_B}
\bm{\mathcal{G}}(ck, \bm{r}-\bm{R})
.
\end{equation}
Further assuming that incident light is near resonant such that $|\frac{ck}{\omega_{\mathrm{ex}}}-1|\ll1$, we find:
\begin{equation}
E_\tau(\bm{r})
\simeq
E_{0,\tau}(\bm{r})
+
\frac{3\pi c\gamma}{\omega_{\mathrm{ex}}}
\sum_{\bm{R}_1,\tau_1}
\sum_{\bm{R}_2,\tau_2}
\mathcal{G}_{\tau,\tau_1}(ck, \bm{r}-\bm{R}_1)
D_{\bm{R}_1,\bm{R}_2}^{\tau_1,\tau_2}(ck)
\bm{E}_{0,\tau_2}(\bm{R}_2)
.
\end{equation}

To proceed, we specifically consider plane-wave incident field $\bm{E}_0(\bm{s}+z\bm{e}_z)=\bm{E}_0e^{i\bm{k}\cdot(\bm{s}+z\bm{e}_z)}$ with $k_z\equiv\bm{k}\cdot\bm{e}_z>0$.
We are particularly interested in the far-field limit $|z|\to\infty$, which allows for the approximation Eq.~\eqref{eq:g_r_ck_kB_b}.
Combining this with the following relations:
\begin{equation}
D_{\bm{L}+\bm{b},\bm{L}'+\bm{b}'}^{\tau,\tau'}(ck)
\equiv
\frac{1}{N}
\sum_{\bm{Q}}
e^{i\bm{Q}\cdot(\bm{L}+\bm{b}-\bm{L}'-\bm{b}')}
D_{\bm{b},\bm{b}'}^{\tau,\tau'}(ck,\bm{Q})
,\quad
\sum_{\bm{L}'}
e^{i(\bm{k}-\bm{Q})\cdot(\bm{L}'+\bm{b}')}
=
N
\delta_{\bm{k}-(\bm{k}\cdot\bm{e}_z)\bm{e}_z,\bm{Q}}
,\quad
N=\sum_{\bm{L}}1
,
\end{equation}
where $\bm{b}$ again denote sublattice vectors for non-Bravais lattices, we find the following scattering formula:
\begin{equation}
E_\tau(\bm{s}+z\bm{e}_z)
\simeq
\left[
E_{0,\tau}
e^{ik_zz}
+
e^{-ik_zz}
\sum_{\tau'}
S_{\tau,\tau'}(\bm{k})
E_{0,\tau'}
\right]
e^{i\bm{k}\cdot\bm{s}}
,\quad
z\to -\infty
.
\end{equation}
Here the (reflection sector) of scattering matrix reads (expressing $\bm{k}=\bm{k}_{||}+k_z\bm{e}_z$):
\begin{equation}
\label{eq:S_matrix}
S_{\tau,\tau_2}(\bm{k})
=
\frac{3\pi c\gamma}{\omega_{\mathrm{ex}}}
\sum_{\tau_1}
g_{\tau,\tau_1}(\bm{k})
D_{\tau_1,\tau_2}(ck,\bm{k}_{||})
,\;
g_{\tau,\tau_1}(\bm{k})
=
\bm{e}_{\tau}^\ast
\cdot
\bm{g}(ck,\bm{k}_{||})
\cdot
\bm{e}_{\tau_1}
,\;
D_{\tau_1,\tau_2}(ck,\bm{k}_{||})
=
\sum_{\bm{b}_1,\bm{b}_2}
D_{\bm{b}_1,\bm{b}_2}^{\tau_1,\tau_2}(ck,\bm{k}_{||})
.
\end{equation}
Note that all sublattices are symmetized in $D_{\tau_1,\tau_2}(ck,\bm{k}_{||})$.
For Bravais lattices, we recover the scattering formula presented in the main text.

\section{Parameters}

In this section, we summarize the parameters discussed  utilized for numerical calculation.
We consider the values from WSe\textsubscript{2}/WS\textsubscript{2}~\cite{park2023dipole}, which gives moir\'e period $a_M = 8.25$nm and exciton Wannier orbital center-of-mass localization length $a_W = 2$nm at zero twist.
At small but finite twisting angle $\theta$, the moir\'e period scale as $a_M(\theta) = a_M(0)\frac{\delta}{\sqrt{\theta^2+\delta^2}}$, where $\delta = 0.04$ is the lattice mismatch, and $a_W(\theta) = a_W(0)\sqrt{\frac{a_M(\theta)}{a_M(0)}}$~\cite{wu2018hubbard}.
Interlayer excitons therein with bare frequency $\omega_{\mathrm{ex}} = \frac{2\pi c}{\lambda_{\mathrm{ex}}} = 1.55$eV are considered.
The exciton Bohr radius $a_B$ and the transition dipole $d$ are packed into the bare radiative decay rate $\gamma$, which acts as an unit for energy variables and therefore is set as one in the numerical computation. 
Finally, Zeeman splitting from the out-of-plane magnetic field are either set as zero or $\mu_BB = 20\gamma$.

\section{Supplementary data for collective excitonic bands}
\label{sec:supplementary_data}

\begin{figure}[h]
    \centering
    \includegraphics[width=\columnwidth]{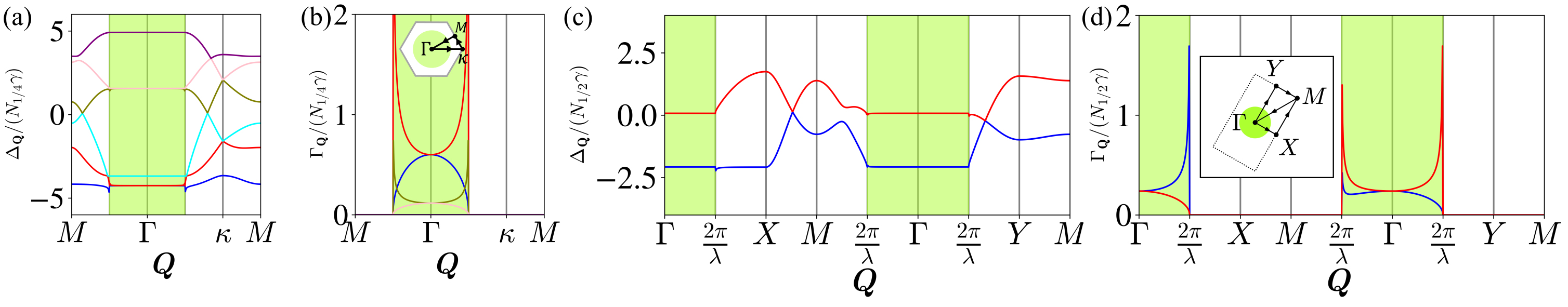}
    \caption{Collective excitonic lineshifts $\Delta_{\bm{Q}}$ and linewidths $\Gamma_{\bm{Q}}$ emerging from charge ordered zero-twist WSe\textsubscript{2}/WS\textsubscript{2} with electron fillings (a,b) $\nu_e=\frac{1}{4}$ and (c,d) $\nu_e=\frac{1}{2}$. 
    The vertical axes are displayed in units of $\gamma N_{\nu_e}$.
    The horizontal axes show Bloch momenta $\bm{Q}$ at high symmetry points, which follow a piecewise-linear path in the Brillouin zone, as depicted by the dashed hexagon and rectangle in the insets of (b) and (d), respectively.
    Momenta within the light cone are indicated by the green shaded area (size enlarged for clarity).
    Different colors label distinct single-particle exciton bands.
    Parameters are chosen to be the same as Fig.~2 of the main text.
    }
    \label{Fig_S1}
\end{figure}

In this section, we present numerical results for various lattices.
As mentioned in the main text, we label the collective bands $\Delta_{\bm{Q}}-\frac{i}{2}\Gamma_{\bm{Q}}$ in energy order as $\Lambda = 0,1,2...$, and scale the eigenvalues by $N_{\nu_e}=
\lambda_{\mathrm{ex}}^2/\mathcal{A}_{\nu_e}$, where $\lambda_{\mathrm{ex}} = \frac{2\pi c}{\omega_{\mathrm{ex}}}$ and $\mathcal{A}_{\nu_e}$ denotes the emergent unit cell area at electron filling $\nu_e$ and zero twist.

Fig.~\ref{Fig_S1}(a) and (b) show the collective excitonic bands from Kagome lattice at $\nu_e=\frac{1}{4}$.
There are six bands due to two coupled valleys and three sublattices within an emergent unit cell~\footnote{Although the number of collective bands should equal two (coupled valleys) times the number of sublattices per unit emergent cell, we note that some states are degenerate and non-resolvable at high symmetry momenta.}.
Four of them exhibit enhanced cooperative decay rate (compared to $\gamma$), and the lowest doublet ($\Lambda=0,1$) possesses a larger $\Gamma_{0}$ than the other one ($\Lambda=3,4$).
Notably, these four bright states are compressed into two valley components of the scattering matrix Eq.~\eqref{eq:S_matrix} upon sublattice symmetrization of the exciton propagator, indicating that $S_{\tau,\tau'}(\bm{k})$ only contains partial information of the excitonic collective states.
In contrast, the states $\Lambda=2,5$ do not show a significant radiative decay rate, and unlike others, they are non-degenerate at $\bm{Q}=0$.

Fig.~\ref{Fig_S1}(c) and (d) demonstrate the exciton eigenvalues from rectangular lattice at $\nu_e=\frac{1}{2}$.
There are two bands due to two coupled valleys.
Both of them exhibit enhanced cooperative decay rate (compared to $\gamma$), which are non-degenerate at $\bm{Q}=0$ unlike the ones from triangular, honeycomb, and kagome lattices.

\begin{figure}[h]
    \centering
    \includegraphics[width=0.5\columnwidth]{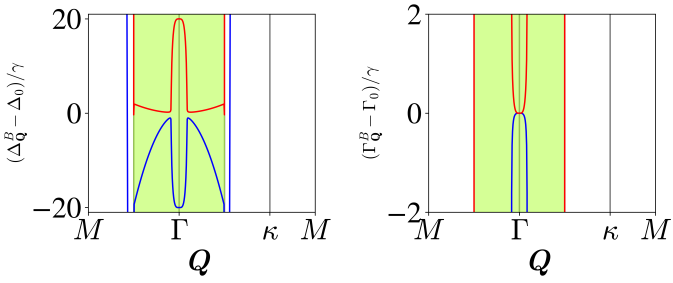}
    \caption{
    Collective excitonic bands in the presence of Zeeman splitting $\mu_BB=20\gamma$, denoted as $\Delta_{\bm{Q}}^B-\frac{i\Gamma_{\bm{Q}}^B}{2}$ for triangular lattice at $\nu_e=0$.
    The vertical axes are shifted with respect to the zero-magnetic-field spectrum at zero momentum $\Delta_0-\frac{i\Gamma_0}{2}$ and are scaled by $\gamma$.
    Other parameters are the same as Fig.~2 of the main text.
    }
    \label{Fig_S2}
\end{figure}

Fig.~\ref{Fig_S2} shows the collective excitonic bands for triangular lattice at $\nu_e=0$ in the presence of out-of-plane magnetic field, which splits the valley degeneracy at $\bm{Q}=0$.
Similar splitting also occurs for honeycomb and Kagome lattices (not shown). In contrast, for rectangular lattice, the two collective bands are non-degenerate even in the absence of magnetic field, indicating that the Zeeman term does not quanlitatively modify the spectrum.

\begin{figure}[h]
    \centering
    \includegraphics[width=\columnwidth]{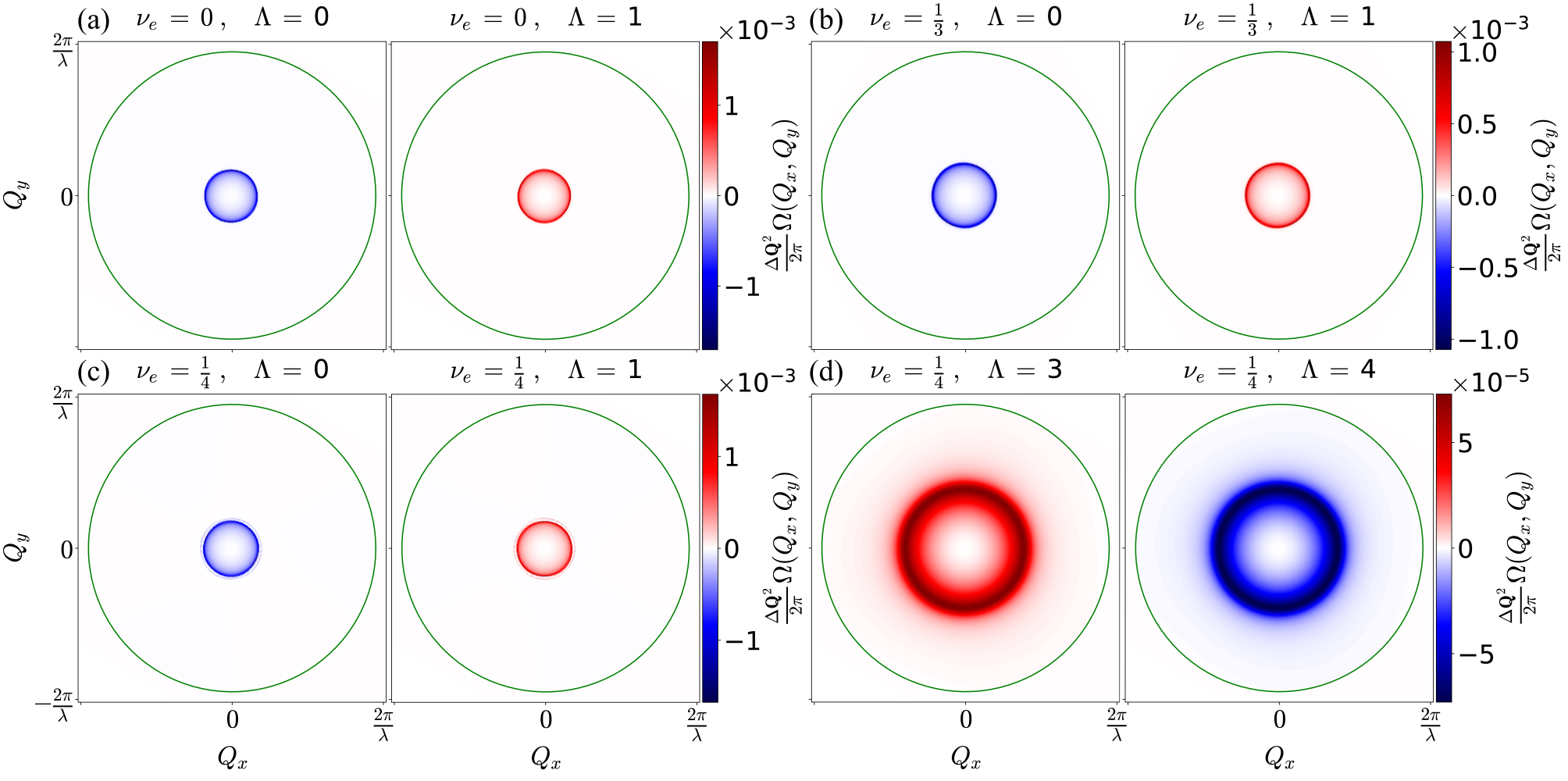}
    \caption{Berry curvatures $\Omega(Q_x,Q_y)$ of collective bands with $\Gamma_{\bm{Q}}\gg\gamma$ from emergent arrays at various electron doping in the presence of Zeeman splitting $\mu_BB=20\gamma$.
    Other parameters are the same as Fig.~2 of the main text.
    }
    \label{Fig_S3}
\end{figure}

The Berry curvatures ($\Omega$) of the states with collectively enhanced radiation from triangular, honeycomb, and Kagome lattices are plotted in Fig.~\ref{Fig_S3}(a), (b), and (c,d), respectively (all the states with $\Gamma_{\bm{Q}}\ll\gamma$ yield suppressed $\Omega$ and hence are not shown).
Note that here we add an out-of-plane magnetic field to split the degeneracies at the Brillouin zone center (such that $\Omega$ is well-defined).
These states generally appear as doublets; each one contains two states with opposite Berry curvatures $\pm\Omega$.
Among these emergent lattice structures, $\Lambda = 0,1$ yield similar $\Omega(\bm{Q})$. 
In contrast, the states $\Lambda = 3,4$ at $\nu_e=\frac{1}{4}$ give much broader distributions of Berry curvature.

\bibliography{Biblio}